\pdfoutput=1

\documentclass[11pt]{article}

\usepackage[pdftex]{graphicx,color} 
\usepackage[utf8]{inputenc}
\usepackage[makeroom]{cancel}	% gamma-tech-stuff
\usepackage[normalem]{ulem}		% underline

\usepackage{jheppub}
\usepackage{multirow}
\usepackage{mathtools}
\usepackage{comment}
\usepackage{dsdshorthand}
\usepackage{framed}
\usepackage{amsmath,amsfonts,amsthm,amssymb}	% Math packages & curly letters
\usepackage{slashed}
\usepackage{multirow}
\usepackage{bbold}
\usepackage{mathbbol}
\usepackage{bbm}
\usepackage{simplewick}		% Wick contractions 
\usepackage{makecell}		% Thicker table row lines. Summon with \xhline{3\arrayrulewidth} .
\usepackage{tabu}			% Thicker table columns lines. Use tabu instead of tabular.

\numberwithin{equation}{section}

\newcommand{\bea}{\begin{equation}\begin{aligned}}
\newcommand{\eea}[1]{\label{#1}\end{aligned}\end{equation}}
\newcommand{\beq}{\begin{equation}}
\newcommand{\eeq}{\end{equation}}
\newcommand{\nn}{\nonumber\\}

\def\D{\Delta}
\def\g{\gamma}
\def\a{\alpha}

\newcommand{\eq}{&\quad}
\newcommand{\tq}{&\quad\quad}
\newcommand{\rig}{\right.}
\newcommand{\lef}{\left.}
\newcommand{\para}{\parallel}
\newcommand{\vev}[1]{\left\langle #1 \right\rangle}

\newcommand{\disc}{\text{disc}}
\newcommand{\Li}{\text{Li}}

\newcommand{\lhs}{\text{LHS}}
\newcommand{\rhs}{\text{RHS}}
\newcommand{\boe}{\text{boe}}
\newcommand{\ope}{\text{ope}}
\newcommand{\fr}{\text{(free)}}

\newcommand{\hD}{{\hat{\Delta}}}
\newcommand{\hO}{{\hat{O}}}
\newcommand{\hp}{{\hat{\phi}}}

\newcommand{\hg}{\hat{\g}}
\newcommand{\hta}{\hat{\tau}}
\newcommand{\hT}{\hat{T}}

\newcommand{\pa}{\partial}
\newcommand{\mco}{\mathcal{O}}

\newcommand{\mR}{\mathbb{R}}

\newcommand{\mG}{\mathcal{G}}
\newcommand{\mcg}{\mathcal{G}}
\newcommand{\1}{\mathbb{1}}

\newcommand{\al}{\alpha}
\newcommand{\bet}{\beta}
\newcommand{\ch}{\chi}
\newcommand{\del}{\delta}

\newcommand{\ph}{\phi}

\newcommand{\kap}{\kappa}
\newcommand{\la}{\lambda}
\newcommand{\m}{\mu}
\newcommand{\n}{\nu}

\newcommand{\si}{\sigma}

\newcommand{\up}{\upsilon}

\newcommand{\Ph}{\Phi}

\usepackage{tikz}
\usetikzlibrary{arrows,calc,shapes,decorations.pathmorphing,positioning,decorations.markings}
\tikzset{
>=stealth',
help lines/.style={dashed, thick},
axis/.style={<->},
important line/.style={thick},
connection/.style={thick, dotted},
  cross/.style={
    cross out,
    draw=black, 
    minimum size=7pt, 
    inner sep=0pt,
    outer sep=0pt
  },
  branchcut/.style={
    decoration={
      snake,
      amplitude=1pt,
      segment length=6pt,
    },
    decorate,
    thick
  },
->-/.style={decoration={
  markings,
  mark=at position #1 with {\arrow{>}}},postaction={decorate}}
}

\makeatletter
\gdef\@fpheader{}
\makeatother
\preprint{UUITP-55/20}

\title{On Analytic Bootstrap for Interface and Boundary CFT}

\author{Parijat Dey and Alexander Söderberg}

\affiliation{Department of Physics and Astronomy,
	Uppsala University,\\
	Box 516,
	SE-751 20 Uppsala,
	Sweden}

\emailAdd{parijat.dey@physics.uu.se}\emailAdd{alexander.soderberg@physics.uu.se}

\abstract{We use analytic bootstrap techniques for a CFT with an interface or a boundary. Exploiting the analytic structure of the bulk and boundary conformal blocks we extract the CFT data. We further constrain the CFT data by applying the equation of motion to the boundary operator expansion. The method presented in this paper is general, and it is illustrated in the context of perturbative Wilson-Fisher theories. In particular, we find constraints on the OPE coefficients for the interface CFT in $4-\e $ dimensions (upto order $\mco(\e^2)$) with $\f^4$-interactions in the bulk. We also compute the corresponding coefficients for the non-unitary $\f^3$-theory in $6-\e$ dimensions in the presence of a conformal boundary equipped with either Dirichlet or Neumann boundary conditions upto order $\mco(\e)$, or an interface upto order $\mco(\sqrt{\e})$.}

\begin{document}
	
\maketitle

\newtheorem{defin}{Definition}
\newtheorem{thm}{Theorem}
\newtheorem{cor}{Corollary}
\newtheorem{pf}{Proof}
\newtheorem{nt}{Note}
\newtheorem{ex}{Example}
\newtheorem{ans}{Ansatz}
\newtheorem{que}{Question}
\newtheorem{ax}{Axiom}

\section{Introduction}

%%%%%%%%%%%%%%%%%%%%%%%%%%%%%%%%%%%%%%%%%%%%%%%%%%%%%%%%%%%%%%%%%%%%%%%%

%Conformal field theories (CFTs) with boundaries or interfaces describe semi-infinite systems with differently-ordered regions separated by boundaries or interfaces.

Conformal field theories (CFTs) with boundaries or interfaces describe semi-infinite systems with differently-ordered regions. Boundaries and interfaces can be treated as codimension one defects. In a boundary CFT (BCFT), there is only a bulk theory on one side of the defect, with an unphysical region on the other side. In an interface CFT (ICFT), there is a bulk theory on each side of the codimension one defect. Interfaces can be realized physically in different ways. E.g. we can consider a quantum field theory (QFT) with two different vacua, with a finite energy barrier (that allows quantum tunneling) between them. This finite energy barrier can be described effectively by an interface, where the bulk theories correspond to the two vacua. Another realization is to probe a CFT with an operator from another CFT. We can then use the operator/state correspondence on this operator, which yields a small sphere around it. This sphere can be mapped to a line using a conformal transformation, which corresponds to an interface communicating with the two CFTs. An especially interesting case is when we probe a free theory with an interacting one, which ends up with the renormalization group (RG) domain wall studied in \cite{Gliozzi:2015qsa}. It has been speculated that such interface can tell us about the RG flow of the theory.

%An interface can effectively describe the tunneling between two vacua in a quantum field theory separated by a finite energy barrier, or the probing of a field from one CFT into another using the operator/state correspondence and a conformal transformation.  

BCFTs and ICFTs have reduced symmetries compared to \textit{homogeneous CFTs} (without a boundary or interface). However, they can still provide information about the bulk CFTs. These theories contain both the bulk operators as well as operators living on the boundary or the interface, and allow CFT techniques to be extended to a larger domain in the space of QFTs. The bulk CFT data, i.e. the spectrum of bulk operators and operator product expansion (OPE) coefficients, is a local property of the bulk CFT and is unaffected by the presence of the boundary or interface. BCFTs (ICFTs) are also characterized by the additional  data: the spectrum of boundary (interface) operators and the boundary (interface)  operator expansion (BOE/IOE) coefficients. Since the conformal symmetry is partially broken due to the presence of the boundary, the bulk operators can have non-vanishing one-point functions as a result of which the non-trivial observables in such systems are the two-point correlation function of bulk operators. These two-point functions can be expanded in  two configurations. One can consider the case where both the operators are close to the boundary but far from each other. This is known as the boundary-channel expansion  which involves the boundary operator dimensions and the BOE coefficients. On the other hand, the two-point function can be expanded in another channel where both the operators are close to each other but far from the boundary. This is known as the bulk-channel expansion and contains the bulk spectrum and OPE coefficients.  A detailed analysis of BCFTs in this context can be found in \cite{McAvity:1995zd}. The equality of the bulk- and boundary-channel results in a bootstrap equation for the BCFT that can be used to study the bulk and boundary data. This was initiated in  \cite{Liendo:2012hy}. Bootstrap techniques for  BCFT and ICFT were studied  in \cite{Gliozzi:2015qsa}. See  \cite{Prochazka:2019fah, BCFT:2005, Dey:2020lwp, BCFT:200615, Giombi:2020xah, BCFT:2009, Gimenez-Grau:2020jvf, Prochazka:2020vog} for recent works on BCFTs.
%\PDnote{Add relevant references .} 

%The perturbative computation of the BCFT/ ICFT data using Feynman diagrams involves many Feynman diagrams at higher loop levels. One can bypass these Feynman diagram computations by the bootstrap techniques to compute the CFT data.

%In \cite{Liendo:2012hy} the BCFT data was computed by bootstrapping the correlator for the Wilson-Fisher (WF) theory in $4-\e$ dimensions upto $\mco(\e)$.
%The perturbative computation 
%The approach in \cite{Bissi:2018mcq} makes use of the analytic structure of the branch cuts in the bulk- and boundary-channel (which simplifies for integer scaling dimensions) to extract the CFT data, although it is more suited towards theories where only even or odd operators (w.r.t. the scaling dimensions at $\mco(\e^0)$) are present in the boundary-channel.
In perturbation theory where we have an expansion of the CFT data in terms of the perturbative parameter, the computation involves many Feynman diagrams at higher loop orders. One can simplify these computations by the bootstrap techniques to compute the BCFT/ICFT data with some inputs from the Feynman diagrams for the bulk CFT. The BCFT data in $4-\e$ dimensions at the Wilson-Fisher (WF) fixed point was computed analytically upto $\mco(\e)$ in \cite{Liendo:2012hy} using conformal bootstrap methods. This was further studied upto $\mco(\e^2)$ in \cite{Bissi:2018mcq} exploiting the analytic properties of the conformal blocks. One can also use analytic functionals to study BCFTs \cite{Kaviraj:2018tfd, Mazac:2018biw}. The approach in \cite{Bissi:2018mcq} makes use of the analytic structure of the branch cuts in the bulk- and boundary-channel to extract the CFT data, although it is more suited towards theories where only even or odd operators (w.r.t. the scaling dimensions at $\mco(\e^0)$) are present in the boundary-channel.  We modify this method such that it yields more constraints on the CFT data in theories where both even and odd operators in the boundary-channel may appear. 

The method we present consists of two parts. First we study the bootstrap equation and its analytic structure to constrain the CFT data. This is done completely without reference to the Lagrangian. In the second part we make use of the Lagrangian description of the system, in particular its equation of motion (e.o.m.), together with the BOE to further constrain the CFT data. This is similar to the method in \cite{Rychkov:2015naa}, which has also been applied to the $\Z_2$- and $O(N)$-twist defect \cite{Yamaguchi:2016pbj, Soderberg:2017oaa}, and more recently to BCFTs \cite{Giombi:2020rmc}. We find that in the free theory of any unitary scalar CFT with a Lagrangian description, only the fundamental scalar and its normal derivative may appear in the BOE. We proceed to consider a BCFT in $3 - \e$ dimensions as well as a BCFT and ICFT in $4 - \e$ dimensions, and expand in $\e$ to find the anomalous dimension for the boundary operators from the free theory. % at the lowest order. Our results coincide with the literature.

The primary model we study is the CFT with an interface near four dimensions. We focus on the two-point correlation function of bulk scalar operators. From the bootstrap equation we impose constraints on the CFT data. We compute the correlator by resumming this data. Then we impose the e.o.m. on the correlator to further constrain the CFT data. The constraints we find are summarised in subsection \ref{CFTdatasummary}. In the limit when one side of the interface is free, we make contact with the RG domain wall \cite{Gliozzi:2015qsa}. Finally we explore  the CFT with a cubic interaction near six dimensions in the presence of a boundary subject to Dirichlet or Neumann boundary conditions (b.c.'s), or an interface. %We have shown how  the equation of motion for these theories can be used to constrain the CFT data further.
%\PDnote{What is the advantage of this approach over others? Motivation}
%The main result of the current paper is given in \eqref{icftdata2}, \eqref{icftdata3}, \eqref{icftdata4},  \eqref{Full Corr} which gives the ICFT data as well as the full correlator $\langle\f \f \rangle$ upto $\mco(\e^2)$ in terms of the bulk and boundary spectrum of the theory.

The paper is organised as follows. 
%We focus on the interface CFTs and boundary CFTs in the Wilson-Fisher theory in $4-\e$ and $6-\e$ dimensions respectively. 
In section \ref{Sec: Blockology} we discuss the analytic structure of the conformal blocks and how these can be used to analyze the bulk and boundary data. We discuss the entire methodology that we use to constrain the CFT data with a discussion on the similarities and differences with \cite{Bissi:2018mcq}. We also find a general expression for the bulk OPE coefficients \eqref{Bulk coeff} in $d$ dimensions at order $\e^k$ in terms of the anomalous dimensions as well as the OPE coefficients from the previous order in the expansion parameter. In section \ref{Sec: Free ICFT} we study the bootstrap constraints on the two-point correlator of scalar operators in $4-\e $ dimensions in an  ICFT with scalars on both sides of the interface, transforming in the fundamental representation of $O(N) \times O(N)$. This is followed by further constraints on the data from the e.o.m.. In section \ref{6d BCFT} we  bootstrap the scalar correlator for $\f^3$-theory in $6-\e$ dimensions in a BCFT with Dirichlet/Neumann b.c.. Section \ref{Sec: 6d ICFT} contains the study of an ICFT in $6-\e$ dimensions. We conclude in section \ref{Conc} with some open questions and future directions. The appendices give the calculational details.

%%%%%%%%%%%%%%%%%%%%%%%%%%%%%%%%%%%%%%%%%%%%%%%%%%%%%%%%%%%%%%%%%%%%%%%%

\section{Analytic structure of the conformal blocks} \label{Sec: Blockology}

In this section we discuss the analytic structure of the conformal blocks in a CFT with a boundary. We consider a BCFT defined in a $d$-dimensional semi-infinite space \\
$\R^d_+= \{x=(x_\para,z): x_\para \in \mR^{d - 1},  z > 0\}$ bounded by a flat $(d - 1)$-dimensional hypersurface at $z=0$. Due to the boundary at $z=0$ the translational invariance along the $z$-direction is broken whereas this invariance is preserved in the $x_\para$-direction.

%We will consider scalar theories with $O(N)$-symmetry, and assume that there are no boundary effects that break this global symmetry. If the $O(N)$-symmetry is explicitly broken (e.g. by  an external field ), which would yield non-zero one-point functions for the scalar, the problem is more complex and requires more advanced methods \cite{Dey:2020lwp, Shpot:2019iwk}. However, the $O(N)$-symmetry could also be broken by quadratic boundary/interface terms in the action. Then the one point function would only be non-zero at the bulk critical point if these interactions favor order at the boundary/ interface and spontaneous symmetry breaking is possible to imply surface order.

We will consider scalar theories with $O(N)$-symmetry, and assume that there are no boundary effects that break this global symmetry (e.g. by  an external field or terms quadratic in the fields on the boundary/interface action). If the $O(N)$-symmetry is explicitly broken, which would yield non-zero one-point functions for the scalar, the problem is more complex and requires more advanced methods \cite{Dey:2020lwp, Shpot:2019iwk}. I.e. if the breaking is due to quadratic boundary-terms, then the bulk one-point function is non-zero at the bulk critical point if these interacting boundary conditions favour order at the boundary/interface.

%If the $O(N)$-symmetry is not broken however, 
The two-point correlation function of two bulk scalar operators $\phi$ with scaling dimension $\D_\f$ is given by
\begin{align}\label{2pt}
\langle \f(x) \,\f(y)\rangle = \frac{F(\xi)}{|x - y|^{2\D_\f}} \,,
\end{align}
where the cross-ratio $\xi$ reads
\begin{align}
 \xi = \frac{s_\para^2 + (z - z')^2}{4z z'} \ , \quad s_\para^a \equiv x_\para^a - y_\para^a \ , \quad a\in\{1, ..., d - 1\} \ . %\xi=\frac{(x-x')^2}{4 z z'} \ ,
\end{align}
%Here $s_\para^a \ , a\in\{1, ..., d - 1\}$ is the difference between the positions along the boundary. 
The two-point function can be decomposed in the bulk- or  boundary-channel %in the two different limits $\xi \rightarrow 0$ and $\xi \rightarrow \infty$ respectively.
\begin{align}\label{Bootstrap eq}
F(\xi)= \sum_{\D \geq 0} \lambda a_{\D}  \mathcal{G}_{\rm{ope}}({\D}; \xi) = \xi^{\D_\f}\sum_{\hat{\D} \geq 0} {\mu}^2_{\hat{\D}} \mathcal{G}_{\rm{boe}}(\hat{\D}; \xi) \ ,
\end{align}
where the conformal blocks are given by \cite{McAvity:1995zd}
\begin{align}\label{block}
\mathcal{G}_{\rm{ope}}({\D}; \xi) &=\xi^{\Delta / 2} {}_2 F_1\left(\tfrac{\De}{2},\tfrac{\De}{2};\De + 1 - \tfrac{d}{2};-\xi\right)\,,\nn
\mathcal{G}_{\rm{boe}}(\hat{\D}; \xi)&=   \xi^{- \hat{\Delta}} {}_2 F_1 \left(\hat{\Delta},\hat{\De} + 1 - \tfrac{d}{2}; 2 \hat{\De} + 2 - d; - \xi^{-1} \right)\,.
\end{align}
In \eqref{Bootstrap eq} the coefficients ${\mu}^2_{\hat{\D}}$ are the BOE coefficients squared and $\lambda a_{\D}$ are the bulk OPE coefficients times the one-point functions. The terms $\D=0$ as well as $\hat{\D}=0$ represent the contribution of bulk and boundary identity operators respectively. 

Let us now look into the analytic structure of the bulk and boundary conformal blocks $\mathcal{G}_{\rm{ope}}$ and $\mathcal{G}_{\rm{boe}}$. Both of the blocks in \eqref{block} have a branch cut at $\xi < 0$ that originates from generic non-integer power of $\xi$. There is an additional branch cut at $\xi < -1$ in the bulk block from the hypergeometric function with argument $-\xi$. The hypergeometric function in the boundary block with argument $-\xi^{-1}$ has a branch cut at $\xi \in (-1, 0)$. This analytic structure has some important consequences  as discussed in \cite{Bissi:2018mcq}. 
Let us define the discontinuity of a function $f(\xi)$ as 
\begin{align}\label{discdef}
\underset{{\xi }}\disc f(\xi) \equiv  \underset{{\a \rightarrow 0^+}}{\rm{lim}} f(\xi+i \a)- f(\xi-i \a)\,.
\end{align}
If we take the discontinuity of the bootstrap equation \eqref{Bootstrap eq} at $\xi < -1$
the boundary blocks with $\hD = \frac{d - 2}{2} + m \ , m\in\Z_{\geq 0}$ disappear from the equation (this holds for any $d$). These boundary operators correspond to normal derivatives, $\pa_\perp^m\hp$, of a scalar with scaling dimension $\frac{d - 2}{2}$. Also, by taking the discontinuity of \eqref{Bootstrap eq} at $-1 < \xi < 0$ we can remove all the bulk blocks with $\D = d - 2 +2 n \ , n\in\Z_{\geq 0}$. Such operators correspond to scalar double traces $\ph\pa^{2n}\ph$ of a scalar with dimension $\frac{d - 2}{2}$.

We can understand the convergence of the bootstrap equation \eqref{Bootstrap eq} under this analytic continuation using the radial coordinates introduced in \cite{Lauria:2017wav} for defect CFTs. There will be one radial coordinate for  each bootstrap-channel. The one for the boundary-channel changes sign under this analytic continuation, and is still within the region of convergence. On the other hand, the radial coordinate in the bulk-channel approaches the boundary under this analytic continuation. This means that we should tread carefully and explicitly check that the bootstrap channels have the assumed branch cuts after resummation of the bulk OPE coefficients \cite{Bissi:2018mcq}.
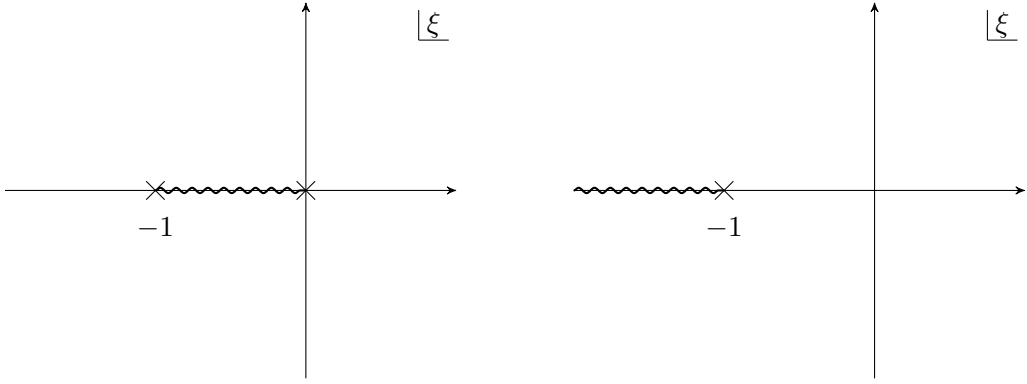
\begin{figure}
\centering
  \begin{tikzpicture}[scale=1]
    \coordinate (n) at (1,2.5);
    \coordinate (e) at (3,0);
    \coordinate (w) at (-3,0);
    \coordinate (s) at (1,-2.5);
    \coordinate (bp1) at (1,0);
    \coordinate (bp2) at (-1,0);
    \draw[->] (w) --  (e) ;
    \draw[->] (s) --  (n) ;
	\draw [branchcut] (bp2) -- (bp1);
    \node at (1,0) [cross] {};
    \node at (-1,0) [cross] {};
    \node at (-1,-0.5) [] {$-1$};
    \node at (2.7,2.2) [] {$\xi$};
    \draw[-] (2.5,2) --  (2.9,2) ;
    \draw[-] (2.5,2) --  (2.5,2.4) ;
  \end{tikzpicture}\qquad \qquad
  \begin{tikzpicture}[scale=1]
    \coordinate (n) at (1,2.5);
    \coordinate (e) at (3,0);
    \coordinate (w) at (-3,0);
    \coordinate (s) at (1,-2.5);
    \coordinate (bp1) at (1,0);
    \coordinate (bp2) at (-1,0);
    \draw[->] (w) --  (e) ;
    \draw[->] (s) --  (n) ;
	\draw [branchcut] (w) -- (bp2);
    \node at (-1,0) [cross] {};
    \node at (-1,-0.5) [] {$-1$};
    \node at (2.7,2.2) [] {$\xi$};
    \draw[-] (2.5,2) --  (2.9,2) ;
    \draw[-] (2.5,2) --  (2.5,2.4) ;
  \end{tikzpicture}
\caption{Analytic structure of  the hypergeometric function in $\mathcal{G}_{\rm{boe}}$ (left) and  $\mathcal{G}_{\rm{ope}}$ (right) in \eqref{block}.}
 \label{fig:analytic_structure_gi_gb}
\end{figure}

The discontinuity of the bulk block at $\xi < -1$  for the bulk exchange   of double-trace operators of dimension $\D = d - 2 + 2n$ can be expressed in terms of  Jacobi polynomials (this holds for any $d$) %\footnote{Similarly one can express $\underset{{\xi \in (-1, 0)}}\disc \mG_{\rm{boe}}(\frac{d - 2}{2} + m ; \xi)$ as a Jacobi polynomial.}
% \footnote{The discontinuity at $\xi=-1$ is proportional to $\delta(\xi+1)$ and one should take that into account when using the orthogonality relation \eqref{Orth Rel}. However this  contribution vanishes due to  the measure factor  $(1+\xi)^{\frac{d}{2}-1}$ in the integrand.}
\begin{align} \label{Disc bulk block}
%\begin{aligned}
\underset{{\xi < -1}}\disc \mG_{\rm{ope}}(d-2+2n ; \xi) &= 2\pi i (-1)^{\frac{d}{2}}\,\xi^{n + \frac{d}{2} - 1}\frac{\G_{2n + \frac{d}{2}-1}}{\G^2_n\,\G_{\frac{d}{2}}}\, _2F_1\left(\frac{d}{2}+n-1,\frac{d}{2}+n-1;\frac{d}{2};\xi +1\right)\nn
& =  -2\pi i (-1)^n \frac{\G_{2n+\frac{d}{2}-1}}{\G_n\,\G_{n-1+\frac{d}{2}}} P_{n - 1}^{(\frac{d}{2}-1, 0)} \left( -\frac{\xi + 2}{\xi} \right)\,,%\nn
% &= -2\pi i\frac{\G({2n + 1})}{\G^2({n})}\xi^{n}\,{}_2F_1(n + 1, n + 1, 2, \xi + 1) \\
% &= \frac{2\pi i}{\xi} \frac{ (-1)^n\G({2n + 1} )}{ \G(n)\G({n + 1}) } P_{n - 1}^{(1, 0)} \left( -\frac{\xi + 2}{\xi} \right) \ .
\end{align}
%\end{equation}
where $\G_x \equiv \G(x)$ is the shorthand notation for the Gamma function and we have used the following identity
\begin{equation}
\begin{aligned}
{}_2F_1(a, b, c, z) &= \frac{1}{(1 - z)^a}\,{}_2F_1\left(a, c - b, c, \frac{z}{z - 1}\right)
\end{aligned}
\end{equation}
in going from the first line to the second line of \eqref{Disc bulk block}. The Jacobi polynomials  satisfy the  orthogonality relation
\begin{equation} \label{Orth Rel}
\begin{aligned}
\int_{-1}^{+1}dy(1 - y)^{\frac{d}{2} - 1}P_{m - 1}^{(\frac{d}{2}-1, 0)}(y)P_{n - 1}^{(\frac{d}{2}-1, 0)}(y) &= \frac{\del_{mn}}{2(m - 1) +\frac{d}{2}} \ ,
%\int_{-\infty}^{-1} \frac{d\xi}{\xi^2}\, {\left(\frac{(\xi+1)}{\xi}\right)}^{\frac{d}{2}-1}\,P_{m-1 }^{(\frac{d}{2}-1, 0)}\left( -\frac{\xi + 2}{\xi} \right)P_{n-1 }^{(\frac{d}{2}-1, 0)}\left( -\frac{\xi + 2}{\xi} \right)&= \frac{\del_{mn}}{2m-2 +\frac{d}{2}} \ , %\quad \forall \ m,n\in\Z_{\geq 1} \ . %\ , \quad \xi = - \frac{2}{y + 1} \ .
\end{aligned}
\end{equation}
which holds for all $m,n\in\Z_{\geq 1}$.

Let us now look into the boundary blocks for the exchange operators of dimensions $\hat{\D}= \frac{d-2}{2} + n$. They satisfy the following orthogonality relation\footnote{This can be found using an ansatz on the form $w^{n - m - 1}{}_2F_1(a_1 + a_2m, a_3 + a_4m, a_5 + a_6m, -w)$ for the orthogonality weight function, where the coefficients $a_i \ , i\in\{1, ..., 6\}$, can be found by studying the residue at $w = 0$ for different values of $m$ and $n$.}
\begin{equation} \label{Boundary Orth Rel}
\begin{aligned}
\underset{|w| = \tilde{\e}}{\oint}\frac{dw}{2\pi i}w^{n - m - 1}{}_2F_1(1 - m, - m - \tfrac{d - 4}{2}, 2(1 - m), -w){}_2F_1(n, n + \tfrac{d - 2}{2}, 2n, -w) = \del_{m,n} \ ,
\end{aligned}
\end{equation}
where $\tilde{\e} \ll 1$. 
%\PDnote{Should we write the orthogonality in terms of $\mathcal{G}_{\rm{boe}}$?}

The boundary conformal blocks have another analytical property that under the transformation $\xi \rightarrow e^{\pm \pi i} (\xi+1)$  they acquire a phase \cite{Bissi:2018mcq}\footnote{This is the same as sending one of the external fields through the boundary $z \rightarrow -z$.}
\begin{equation} 
\begin{aligned}
%\left. \frac{\mG_{\rm{boe}}(\hD; \xi)}{\xi^{\De_\ph}} \right|_{\xi \rightarrow e^{\pm\pi i}(\xi + 1)} = e^{\mp\pi i\hD}\frac{\mG_{\rm{boe}}(\hD; \xi)}{\xi^{\De_\ph}} \ .
\mG_{\rm{boe}}(\hD; e^{\pm\pi i}(\xi + 1)) = e^{\mp\pi i\hD}\mG_{\rm{boe}}(\hD; \xi) \ . %\quad\Rightarrow\quad \mG_{\rm{boe}}(m + 1; -\xi - 1) = (-1)^{m + 1}\mG_{\rm{boe}}(m + 1; \xi) \ .
\end{aligned}
\end{equation}
In particular this means that for integer boundary dimensions $\hat{\D}=m$
\begin{equation} \label{Image symm}
\begin{aligned}
%\left. \frac{\mG_{\rm{boe}}(m; \xi)}{\xi^{\De_\ph}} \right|_{\xi \rightarrow -\xi - 1)} = (-1)^m\frac{\mG_{\rm{boe}}(m; \xi)}{\xi^{\De_\ph}} \ .
\mG_{\rm{boe}}(m ; -\xi - 1) = (-1)^{m }\mG_{\rm{boe}}(m ; \xi) \ .
\end{aligned}
\end{equation}
%We will exploit \eqref{Disc bulk block},\eqref{Boundary Orth Rel} and \eqref{Image symm} to extract the CFT data in Section \ref{Sec: Free ICFT} and \ref{6d BCFT}.

%%%%%%%%%%%%%%%%%%%%%%%%%%%%%%%%%%%%%%%%%%%%%%%%%%%%%%%%%%%%%%%%%%%%%%%%%%%%%%%

%\newpage

\subsection{Methodology}\label{models}

In this subsection we discuss the general method adapted in this paper to extract the CFT data in section \ref{Sec: Free ICFT}, \ref{6d BCFT} and \ref{Sec: 6d ICFT}. %using \eqref{Disc bulk block},\eqref{Boundary Orth Rel} and \eqref{Image symm} . 
Throughout this paper we assume there exist a fundamental scalar, and that the exchanged operators are the scalar double traces $\ph\pa^{2n}\ph$ in the bulk and the normal derivatives $\pa_\perp^m\hp$ on the boundary. That is, in the free theory we have\footnote{This assumption on the exchanged scaling dimensions is motivated from the block decomposition of a generalized free scalar \cite{Liendo:2012hy}.}
%their scaling dimensions are given by
\begin{equation} \label{Scaling dim}
\begin{aligned}
\De_{n \geq 0}^{\fr} &= 2(\De_\ph^{\fr} + n) \ , \quad \hD_{m \geq 0}^{\fr} = \De_\ph^{\fr} + m \ , \quad \De_\ph^{\fr} = \frac{d - 2}{2} \ .
\end{aligned}
\end{equation}
Moreover, we assume a two-point function and its corresponding CFT data is known (up to a finite set of free parameters) at some order in the expansion parameter, say $\mco(\e^{k - 1})$ for some $k \geq 1$.\footnote{E.g. in \cite{Liendo:2012hy} the CFT data was fixed up to one free parameter at $\mco(\e)$ (without using information of the bulk theory as input).} We will denote operators multiplying the $\e^k$-corrections of the OPE coefficients as \textbf{new operators}. Their anomalous dimensions contribute at order $\mco(\e^{k + 1})$. Thus at order $\mco(\e^k)$ it is enough to only consider their free scaling dimensions. If we assume these to be the ones above (for high enough $k$ this will generally not be true, and other operators will be exchanged as well), then their corresponding blocks satisfy the technology discussed in the previous subsection. Using the properties of new operators, the CFT data at $\mco(\e^k)$ can be found using our method.

The method we present is two-fold. Firstly we use the bootstrap equation perturbatively at each order in an expansion parameter to impose constraints on the CFT data.\footnote{Such parameter could e.g. be the deviation from the integer spacetime dimension $\e \ll 1$ or $N^{-1} \ll 1$ in $O(N)$-models. The former will be considered in this paper.} This is done completely without reference to a Lagrangian. Secondly, if we have a Lagrangian description of our system, we impose the e.o.m. on the correlator found from the bootstrap equation. This allows us to further constrain the CFT data of the model.We will outline the essential steps of the method below:

%Let us assume the correlator and its CFT data is known (up to a finite set of free parameters) at some order in the expansion parameter, say $\mco(\e^{k - 1})$ for some $k \geq 1$.\footnote{E.g. in \cite{Liendo:2012hy} the CFT data was fixed up to one free parameter at $\mco(\e)$ (without using information of the bulk theory as input).} The method we present allows us to find the CFT data at $\mco(\e^k)$. We will denote operators multiplying the $\e^k$-corrections of the OPE coefficients as \textbf{new operators}. If we include their anomalous dimensions, their contribution would be $\mco(\e^{k + 1})$. This means that their corresponding blocks satisfy the technology discussed in the previous subsection assuming the exchanged operators have the scaling dimension above. Using their properties, the CFT data at $\mco(\e^k)$ can be found in the following way:

%\newpage

\begin{enumerate}
	\item \label{1} Finding constraints on the CFT data using the bootstrap method:
	\begin{enumerate}
		\item \label{1a} Study the discontinuity of the bootstrap equation along $\xi < -1$, and assume that the discontinuity commutes with the series' of new operators. Due to the analytical structure of the conformal blocks, this assumes that the contribution from new operators in the bulk and boundary have a branch cut along $\xi < -1$ and $\xi \in (-1, 0)$ respectively. The contributions from new bulk OPE coefficients will be given in terms of the Jacobi polynomials \eqref{Disc bulk block}. Using the orthogonality relation \eqref{Orth Rel} for these polynomials we are then able to project out the bulk OPE coefficients. These will be given in terms of the anomalous dimensions of the exchanged operators at the previous order in the expansion parameter. 
		
		%\quad\quad The assumption that operators of scaling dimension $d - 2 + 2n$ and $\frac{d - 2}{2} + m$ are exchanged in the bulk- and boundary-channel respectively is motivated by the conformal block decomposition of a generalized free scalar theory \cite{Liendo:2012hy}.
		% (the block for the "\textit{lowest}" exchanged bulk operator, w.r.t. scaling dimension, does not have a branch cut along $\xi < -1$)
		
		\item \label{1b} Resum the bulk-channel and find the contribution from new operators on the boundary. As was discussed in the previous section, the resummed bulk-channel should only have a branch cut along $\xi < -1$. Regarding the contribution from new operators on the boundary, since we assumed that we could commute the discontinuity with the series' in step \ref{1a}, this contribution is not allowed to have branch cuts along $\xi < -1$. This means that we should set such terms to zero, which constrains the CFT data.
		
		% That is, we study the boundary limit of the bulk field $\ph$ or $\pa_\perp\ph$ and set the term accompanying the BOE coefficient w.r.t. the exchange of $\hp$ or $\pa_\perp\hp$ to zero.
		
		\item \label{1c} If it is a BCFT, we can impose the b.c., e.g. Dirichlet or Neumann, to further constrain the CFT data. In particular, for Neumann b.c.'s when we study the boundary limit of $\pa_\perp\ph$ there may be a pole in $z$ which corresponds to the BOE exchange of $\hp$ which we cannot set to zero. Instead the $z^0$-term should be set to zero, which is the contribution from the boundary operator $\pa_\perp\hp$.
		
		\item \label{1d} We can project out BOE coefficients from the bootstrap equation using the orthogonality relation \eqref{Boundary Orth Rel} for the boundary blocks.\footnote{Alternatively, we may decompose the contribution from new boundary operators in conformal blocks.}
		
		\item \label{1e} The orthogonality relation in the previous step does not state which operators should appear in the BOE.  We thus have to resum the new boundary blocks and check that it reproduces the original contribution. 
		
		\quad\quad \textbf{Even} and \textbf{odd operators} (w.r.t. the boundary scaling dimensions at the lowest order in the expansion parameter) should be resummed separately. These contributions are even or odd respectively under the transformation \eqref{Image symm}, which allows us to find more constraints on the CFT data. Using these constraints, the original contribution from new boundary blocks should equal the contributions from both even and odd operators. %If it does not, fake operators are included. One then has to remove some boundary blocks manually until they are the same.
	\end{enumerate}

	\item Finding constraints on the CFT data using the e.o.m.:
	\begin{enumerate}
		\item Impose the e.o.m. on the correlator found from step \ref{1}. This further constrains the CFT data. However, one has to be a bit careful since a problem with mixing may occur at higher orders in the expansion parameter.
	\end{enumerate}
\end{enumerate}
The $\e$-expansion of a BCFT in $4 - \e$ dimensions (with either Dirichlet or Neumann b.c.'s) was bootstrapped (upto order $\mco(\e^2)$) in \cite{Bissi:2018mcq}. The bootstrap method from that work differs slightly from the one presented here. Step \ref{1b} and \ref{1c} were skipped completely, and only even (Neumann) or odd (Dirichlet) operators were assumed to be exchanged in the BOE. This means that we can use the transformation property (then called \textit{image symmetry}) from step \ref{1e} to constrain the CFT data before finding the BOE coefficients in step \ref{1d}. This simplifies the calculations, although one cannot do this for theories where both even and odd operators appear. In this work we only consider such theories, where our method allows us to constrain the CFT data further.

The bootstrap method presented in this paper reproduces the result from \cite{Bissi:2018mcq} for Neumann b.c.'s. For Dirichlet b.c.'s, the transformation property in step \ref{1e} is trivially satisfied and hence we do not find as many constraints on the CFT data. This means that the method in \cite{Bissi:2018mcq} works better for theories where only even or odd operators appear, but not in the case when both even and odd operators appear.
%In the Dirichlet case there will be fake operators with even scaling dimensions. %The method presented in this work further constrains BCFTs and ICFTs where both even and odd operators appear. We will apply it to the models which contain even and odd operators.
\begin{comment}
The $\e$-expansion of a BCFT in $4 - \e$ dimensions (with either Dirichlet or Neumann BCs) was bootstrapped (up to order $\e^2$) in \cite{Bissi:2018mcq}. The bootstrap method from that work differs slightly from the one presented here in step 1: instead of using step 1b and step 1c, one instead imposes the image symmetry on the boundary blocks (step 1e). This is possible since in four dimensions only even (Neumann) or odd (Dirichlet) operators enter in the BOE. The bootstrap method in step 1 reproduces the result from \cite{Bissi:2018mcq} regarding Neumann BCs. For Dirichlet BCs we get one extra free parameter, although the results are still consistent with \cite{Bissi:2018mcq}. The advantage of step 1 is that it does not assume only even or odd operators in the BOE.
\end{comment}

Since both the discontinuity of the bulk block \eqref{Disc bulk block} as well as the orthogonality relation for the boundary block \eqref{Boundary Orth Rel} hold in any spacetime dimension $d$, the method we present is independent of this parameter. This means that it can also be applied to e.g. three-dimensional theories where the scaling dimensions are half-integers at the lowest order in the expansion parameter.

%If the contribution from new operators does not have integer scaling dimensions at the lowest order in the expansion parameter, its corresponding conformal blocks will have a branch cut along the entire negative real axis $\xi < 0$. In such case, we cannot directly apply this method to bootstrap the two-point correlators. 

If there are infinitely many operators at the previous order in the expansion parameter, we need to use the bulk and boundary anomalous dimensions as input and resum their contribution before we can apply the method presented in this paper. In such case one also has to solve a mixing problem.

\subsection{Bulk OPE coefficients}

Here we will perform step \ref{1a} of the method presented in the previous subsection. This can be done generally for any spacetime dimension if we assume that the operators with scaling dimensions \eqref{Scaling dim} are exchanged upto $\mco(\e^k)$. To illustrate this, assume that we know the correlator and its corresponding CFT data at order $\e^{k - 1}$ for some $k\in\Z_{\geq 1}$ (up to possibly some free parameters).

%Here we will perform step \ref{1a} of the method presented in the previous subsection. This can be done generally for any spacetime dimension if we know the CFT data at the previous order and if we assume that the operators with scaling dimensions \eqref{Scaling dim} are exchanged upto $\mco(\e^k)$.\footnote{Most likely this is not true at all order in $\e$.} To illustrate this, assume that we know the correlator and its corresponding CFT data at order $\e^{k - 1}$ for some $k\in\Z_{\geq 0}$ (up to possibly some free parameters).

%Scaling dimensions with a superscript $\fr$ denotes the ones exchanged in the free theory \eqref{Scaling dim} without anomalous dimensions.

Scaling dimensions with a superscript $\fr$ denotes the ones in \eqref{Scaling dim} without anomalous dimensions. If they do not have this subscript, their anomalous dimensions upto $\mco(\e^k)$ is included as well. We expand the OPE coefficients in the following way\footnote{In section \ref{6d BCFT} we find that near six dimensions we should assume a bulk $\ph$-exchange (instead of $\ph^2$) due to the e.o.m.. This will not affect the formula we find for the bulk OPE coefficients.}
\begin{equation} \label{Exp 1}
\begin{aligned}
\la a_\1 &= \widetilde{\la a}_\1 + \mco(\e^{k + 1}) \ , \\ 
\m_\1^2 &= \tilde{\m}_\1^2 + \mco(\e^{k + 1}) \ , \\
\la a_{\ph^2} &= \widetilde{\la a}_0 + \mco(\e^{k + 1}) \ ,
\end{aligned}
\end{equation}

\begin{equation} \label{Exp 2}
\begin{aligned}
\la a_{\ph\pa^{2n}\ph} &= \widetilde{\la a}_{n} + \e^{k}\la a_n^{(k)} + \mco(\e^{k + 1}) \ , \quad &\text{if } &n\in\Z_{[0, \tilde{n}]} \ , \\
\la a_{\ph\pa^{2n}\ph} &= \e^{k}\la a_n^{(k)} + \mco(\e^{k + 1}) \ , \quad &\text{if } &n\in\Z_{ >\tilde{n} } \ , \\
\m_{\pa_\perp^m\ph}^2 &= \tilde{\m}_{m}^2 + \e^k\m_m^{(k)} + \mco(\e^{k + 1}) \ , \quad &\text{if } &m\in\Z_{[0, \tilde{m}]} \ , \\
\m_{\pa_\perp^m\ph}^2 &= \e^k\m_m^{(k)} + \mco(\e^{k + 1}) \ , \quad &\text{if } &m\in\Z_{ > \tilde{m} } \ .
\end{aligned}
\end{equation}

\noindent Here the CFT data with a tilde is that from order $\e^{k - 1}$, and $\tilde{n}, \tilde{m}\in\Z_{\geq 0}$ denotes the number of exchanged operators at order $\e^{k - 1}$. If infinitely many operators ($\tilde{m}, \tilde{n} \rightarrow +\infty$) are exchanged at order $\e^{k - 1}$, we need to resum their contribution to the bootstrap equation before we can find the OPE coefficients at order $\e^k$. This requires all of the corresponding anomalous dimensions as input.

We will write the bootstrap equation in terms of new operators (these are included in $H_b$ and $H_i$)
\begin{equation} \label{Bootstrap eq 2}
\begin{aligned}
F(\xi) &= G_b(\xi) + H_b(\xi) + \mco(\e^{k + 1}) = G_i(\xi) + H_i(\xi) + \mco(\e^{k + 1}) \ ,
\end{aligned}
\end{equation}
\begin{equation} \label{Gb Gi Hb Hi}
\begin{aligned}
G_b &= \widetilde{\la a}_\1 + \sum_{n = 0}^{\tilde{n}}\widetilde{\la a}_n\mcg_{\ope}(\De_n; \xi) \ , \quad &H_b &= \e^k\sum_{n \geq 1}\la a_n^{(k)}\mcg_{\ope}(\De_n^{\fr}; \xi) \ , \\
G_i &= \tilde{\m}_\1^2\xi^{\De_\ph} + \sum_{m = 0}^{\tilde{m}}\tilde{\m}_m^2\xi^{\De_\ph}\mcg_{\boe}(\hD_m; \xi) \ , \quad &H_i &= \e^k\sum_{m \geq 0}\m_m^{(k)}\xi^{\De_\ph^{\fr}}\mcg_{\boe}(\hD_m^{\fr}; \xi) \ ,
\end{aligned}
\end{equation}

\noindent where $\De_n$ and $\hD_m$ are the full scaling dimensions of $\ph\pa^{2n}\ph$ and $\pa_\perp^m\hp$ respectively, including their anomalous dimensions. Since the orthogonality relation \eqref{Orth Rel} does not hold for the bulk $\ph^2$-exchange, we let its corresponding block be in $G_b$. If we consider the discontinuity along $\xi < -1$ \eqref{Disc bulk block}, and assume that it commute with the series in $H_b$ and $H_i$ we find
\begin{equation} \label{Discs}
\begin{aligned}
\disc_{\xi < -1}H_b &= \e^k\sum_{n\geq 1}\la a_n^{(k)}\disc_{\xi < -1}\mcg_{\ope}(\De_n^{\fr}; \xi) \\
&= -2\pi i\e^k \sum_{n\geq 1} \la a_n^{(k)}(-1)^n \frac{\G_{2n+\frac{d}{2}-1}}{\G_n\,\G_{n + \frac{d}{2} - 1}} P_{n - 1}^{(\frac{d}{2}-1, 0)} \left( -\frac{\xi + 2}{\xi} \right) \ , \\
\disc_{\xi < -1}H_i &= \e^k\sum_{m\geq 0}\m_m^{(k)}\disc_{\xi < -1}\mcg_{\boe}(\hD_m^{\fr}; \xi) = 0 \ .
\end{aligned}
\end{equation}

\noindent Using the orthogonality relation \eqref{Orth Rel}, we find the bulk OPE coefficients as an integral using the bootstrap equation \eqref{Bootstrap eq 2}
\begin{equation} \label{Bulk coeff}
\boxed{
	\begin{aligned}
	\la a_n^{(k)} &= -\frac{ \G_n\G_{n + \frac{d}{2} - 1} }{ \pi i\e^k 2^{\frac{d}{2} + 1}(-1)^n\G_{2n + \frac{d}{2} - 2} } \int_{-1}^{+1}dy \left. (1 - y)^{\frac{d}{2} - 1}P_{n - 1}^{(\frac{d}{2}-1, 0)}(y)\disc_{\xi < - 1} (G_i - G_b) \right|_{\xi = - \frac{2}{y + 1}} \ .
	\end{aligned}
}
\end{equation}

\noindent This formula holds for any spacetime dimension assuming the exchanged operators are $\ph\pa^{2n}\ph$ and $\pa_\perp^m\hp$. $G_i - G_b$ is the theory-dependent part. The rest of the formula is universal. The special case of this formula when $d = 4$ was studied in \cite{Bissi:2018mcq} at order $\e^2$.

%%%%%%%%%%%%%%%%%%%%%%%%%%%%%%%%%%%%%%%%%%%%%%%%%%%%%%%%%%%%%%%%%%%%%%%%%%%%%%%

\section{An interface CFT in $4-\e$ dimensions} \label{Sec: Free ICFT}

In this section we will consider a flat conformal interface spanned along $\R^{d - 1}$ in a CFT in flat space $\R^d$ with a $\phi^4$ bulk interaction in $d = 4-\e$ dimensions, and study the expansion in $\e$. Let us place the interface at the coordinate $z = 0$. The ICFT we consider has an $O(N)\times O(N)$ global symmetry which comes from two scalars $\ph^i_\pm \ , i\in\{1, ..., N\}$, transforming in the fundamental representation of $O(N)$, defined on each side of the interface: \\
$\R^d_+ = \{(x_\para, z): x_\para\in\R^{d -1}, z > 0\}$ and $\R^d_- = \{(x_\para, z): x_\para\in\R^{d -1}, z < 0\}$. On the interface we identify the scalars and their normal derivatives with each other\footnote{See appendix \ref{App: BC} for details regarding this b.c.}
%The homogenous CFT (without the interface) enjoys a $SO(d + 1, 1)$ conformal symmetry in Euclidean space, and if the interface is conformal it preserves the $SO(d, 1)$ conformal symmetry along the interface. 
\begin{equation} \label{Boundary conditions}
\begin{aligned}
\hp_-^i &= \hp_+^i \equiv \hp^i \ , \quad \pa_\perp\hp^i_- = \pa_\perp\hp^i_+ \equiv \pa_\perp\hp^i \ . 
\end{aligned}
\end{equation}
Hatted operators are local fields on the interface that appear in the BOE/IOE of bulk fields. Even in the free theory there may be poles in these expansions coming from the interface limit, see e.g. \cite{Prochazka:2019fah} for a discussion on how the BOE describes mixing of renormalized fields in the boundary limit. 

Above b.c. was also considered in \cite{Herzog:2017xha}, and it reproduces the correlators from \cite{Gliozzi:2015qsa} upto $\mco(\e)$ when one side of the interface is free. Interfaces with different b.c.'s were studied in \cite{BM77, Sym81, BE81, EB82, DDE83}. In particular, they considered either Dirichlet or Neumann b.c.'s on each side of the interface (which were allowed to be different on the two sides of the interface). Since the b.c.'s determine the correlators, we can use them to classify the interface. This means that in this paper we consider another interface, and thus we expect the CFT data we find to differ from their results. 

%In particular, we work near the fixed point $c=0$ of the model (with a $c\hp^2$-term on the interface) studied in \cite{BM77}.\footnote{In \cite{BM77} they consider the strict $N\rightarrow\infty$ limit} Note that in some cases, e.g. the large N expansion, the CFT data can depend on the value of $c$ \cite{EB82}. Though this is not the case for the $\e$-expansion that we consider.

The \textbf{folding trick} "converts" an ICFT to a BCFT simply by shifting $z\rightarrow -z$ in $\hp^i_-$ \cite{WONG1994403}
\begin{equation}
\begin{aligned}
\hp_-^i(x_\para, -z) - \hp^i_+(x_\para, z) &= 0 \ , \quad \pa_\perp\hp^i_-(x_\para, -z) + \pa_\perp\hp^i_+(x_\para, z) = 0 \ . 
\end{aligned}
\end{equation}
These are Neumann ("$+$") and Dirichlet ("$-$") b.c.'s for the linear combinations $\hp^i_+ \pm \hp^i_-$. Let us define a scalar in the representation $(N \times 1) \oplus (1 \times N)$ of $O(N)\times O(N)$
\begin{equation}
\begin{aligned}
\Ph^i_\al(x_\para, z) &= \ph^i_-(x_\para, -z)\del_{\al -} + \ph^i_+(x_\para, z)\del_{\al +} \ , \quad \al=\pm \quad\Leftrightarrow\quad \vec{\Ph} = \begin{pmatrix}
\vec{\ph}_- \\
\vec{\ph}_+
\end{pmatrix} \ ,
\end{aligned}
\end{equation}
and also define projectors $\Pi_\pm$ that project out the linear combinations that satisfy Dirichlet and Neumann b.c.'s
\begin{equation} %\label{Def Proj}
\begin{aligned}
(\Pi_\pm)^{ij}_{\al\bet} &= \frac{\del^{ij}}{2}\begin{pmatrix}
1 & \pm 1 \\
\pm 1 & 1
\end{pmatrix}_{\al\bet} \ , \quad \Pi_\pm\vec{\Ph} = \begin{pmatrix}
\vec{\ph}_- \pm \vec{\ph}_+ \\
\pm\vec{\ph}_- + \vec{\ph}_+
\end{pmatrix} \ .
\end{aligned}
\end{equation}
We find the two-point correlator of $\Ph^i_\al$ in the free theory using the \textbf{method of images}
\begin{equation} \label{Corr}
\begin{aligned}
\vev{\Ph^i_\al(x_\para, z)\Ph^j_\bet(y_\para, z')} &= \vev{\Ph^i_\al(x_\para, z)\Ph^j_\bet(y_\para, z')}_H + \\
\eq + \ch_{\bet\g}^{jk}\vev{\Ph^i_\al(x_\para, z)\Ph^k_\g(y_\para, -z')}_H \ , \\
\vev{\Ph^i_\al(x_\para, z)\Ph^j_\bet(y_\para, z')}_H &= \frac{\del^{ij}\del_{\al\bet}}{\left[ (x_\para - y_\para)^2 + (z - z')^2 \right]^{(\D_\al + \D_\bet)/2}} \ .
\end{aligned}
\end{equation}
Here $\D_\al$ is the scaling dimension of $\ph_\al^i$, the subscript $H$ denotes that it is the correlator in a homogeneous CFT, and $\ch^{ij}_{\al\bet}$ is the difference between the projectors, i.e. the difference between fields satisfying Neumann and Dirichlet b.c.'s
\begin{equation} %\label{Corr Prop}
\begin{aligned}
\ch^{ij}_{\al\bet} &= (\Pi_+)^{ij}_{\al\bet} - (\Pi_-)^{ij}_{\al\bet} = \del^{ij}\begin{pmatrix}
0 & \ \ 1 \\
1 & \ \ 0
\end{pmatrix}_{\al\bet} \ .
\end{aligned}
\end{equation}
The correlator \eqref{Corr} can be written in terms of the cross-ratio $\xi$ in \eqref{2pt}
\begin{equation} \label{Corr 2}
\begin{aligned}
\vev{\Ph^i_\al(x)\Ph^j_\bet(y)} &= \frac{\del^{ij}F_{\al\bet}(\up)}{|x - y|^{\D_\al + \D_\bet}} \ , \quad F_{\al\bet}(\up) = \begin{pmatrix}
1 & \up^{\D_\al + \D_\bet} \\
\up^{\D_\al + \D_\bet} & 1 \\
\end{pmatrix}_{\al\bet}  \ ,
\end{aligned}
\end{equation}
where $\up$ is defined as
\begin{equation}
\begin{aligned}
\up^2 &= \frac{\xi}{\xi + 1} \ .
\end{aligned}
\end{equation}
The two-point functions in a BCFT with Neumann or Dirichlet b.c.'s are superpositions of these ICFT correlators: $F_{\text{BCFT}}^\pm = F_{++} \pm F_{+-}$.\footnote{Using the results of this paper and \cite{Bissi:2018mcq}, one can check that this holds upto $\mco(\e^2)$.} Note that $F_{\pm\pm}$ corresponds to the $\ph_\pm-\ph_\pm$ two-point functions with fields on the same side of the interface, and that $F_{+-} = F_{-+}$ corresponds to the $\ph_+-\ph_-$ two-point functions with fields on opposite sides of the interface. Since $\ph_+$ and $\ph_-$ are on different sides of the interface, there is no well defined bulk OPE between them. This means that $F_{+-}$ does not have a well-defined bulk-channel, but only a boundary-channel decomposition. Only $F_{\pm\pm}$ satisfy the bootstrap equation \eqref{Bootstrap eq}. 

Expressed in terms of $x$ and $y$, this correlator is on the same form ($|x - y|^{-(\D_\al + \D_\bet)}$) as a two-point correlator in a homogeneous CFT, which is a consequence of the b.c.'s \ref{Boundary conditions}. At higher orders in the expansion parameter it will differ though.

We will specialize to the case when the two external operators are the fundamental scalar
\begin{equation} \label{Free scaling dim}
\begin{aligned}
\D_+ = \D_- = \D_\ph^{(free)} &= \frac{d - 2}{2} \ .
\end{aligned}
\end{equation}
The free theory decomposition is then
\begin{equation} \label{Free Thy Decomp}
\begin{aligned}
\la a_\1 &= 1 \ , \quad &\la a_\D &= 0 \ , \\ 
\m_\1^2 &= 0 \ , \quad &\m^2_\hD &= \del_{\hD, \D_\ph} + \frac{\D_\ph}{2}\del_{\hD, \D_\ph + 1} \ , \\ 
(\m_\1^2)_\pm &= 0 \ , \quad &(\m^2_\hD)_\pm &= \del_{\hD, \D_\ph} - \frac{\D_\ph}{2}\del_{\hD, \D_\ph + 1} \ .
\end{aligned}
\end{equation}
The interface operators with scaling dimensions $\D_\f$  and   $\D_\f + 1$ correspond to $\hp$ and $\pa_\perp\hp$ respectively. We labelled the IOE coefficients from $F_{+-}$ with a "$\pm$" subscript.

\subsection{Constraints from the bootstrap equation} \label{Sec: ICFT Bootstrap}

In this subsection we will find constraints on the CFT data by solving the bootstrap equation \eqref{Bootstrap eq} in $d = 4 - \e$ dimensions. We will consider ($\ph^4$-)interactions only in the bulk, and assume that they are the same on both sides of the interface. This means that $F_{++}$ is the same as $F_{--}$ at all orders in perturbation theory. We will assume that the coupling constants are proportional to $\e$ at their RG fixed point. 

We begin by expanding the CFT data in $\e$
\begin{equation}
\begin{aligned}
\D_\ph &= \D_\ph^{(free)} + \e\g_\ph^{(1)} +\e^2 \g_\ph^{(2)} + \mco(\e^3) \ ,
\end{aligned}
\end{equation}
\begin{equation}
\begin{aligned}
\D_{n\geq 0} &= 2(\D_\ph^{(free)} + n) + \e\g_n^{(1)} + \e^2\g_n^{(2)} + \mco(\e^3) \ , \\
\hD_{m\geq 2} &= \D_\ph^{(free)} + m + \e\hg_m^{(1)} + \e^2\hg_m^{(2)} + \mco(\e^3) \ , \\
\end{aligned}
\end{equation}

\begin{equation}
\begin{aligned}
\la a_\1 &= 1 \ , \quad &\la a_{\D_n} &= \e\la a_n^{(1)}+\e^2 \la a_n^{(2)}+ \mco(\e^3) \ , \\ 
\m_\1^2 &= 0 \ , \quad &\m^2_{\hD_m} &= \del_{m 0} + \frac{\D_\ph^{(free)}}{2}\del_{m 1} + \e\m_m^{(1)} +\e^2\m_m^{(2)} + \mco(\e^3) \ .
\end{aligned}
\end{equation}
Here $\D_\ph^{(free)}$ is given by \eqref{Free scaling dim}. One can bootstrap the order $\e$ terms by expanding in $\xi$ around zero which  %while assuming that there are no new operators on the interface.
%\begin{equation}
%\begin{aligned}
%\m_{\geq 2}^{(1)} &= 0 \ .
%\end{aligned}
%\end{equation}
allows us to fix the CFT data up to two  parameters $\al$ and $\bet$
\begin{equation} \label{Anom Dim}
\begin{aligned}
\g_\ph^{(1)} &= 0 \ , \quad &\hg_{1}^{(1)} = \hg_{0}^{(1)} &= -\al \ , \\
\la a_n^{(1)} &= (\al + \bet)\del_{n, 0} + \frac{\al}{2}\del_{n, 1} \ , \quad &\m_m^{(1)} &= \bet\del_{m, 0} + \frac{\al - \bet}{2} \del_{m, 1} \,.%\quad \m_{\geq 2}^{(1)} &= 0\ .
\end{aligned}
\end{equation}
These constraints are similar to those found when bootstrapping a BCFT correlator at order $\e$ \cite{Liendo:2012hy}, with one major difference: in this case $\al$ is not related to any of the bulk anomalous dimensions, i.e. we cannot use bulk CFT data as input to find the OPE coefficients and the anomalous dimensions of boundary fields. Moreover there is one more free parameter in the ICFT case.

Let us proceed with bootstrapping the correlator at order $\e^2$ using the method described in section \ref{Sec: Blockology}.\footnote{We cannot naively expand in $\xi$ since only the interface-channel will then contain $\log(\xi)^2$-terms.} The boundary conformal blocks $\mG_{\rm{boe}}(\hD_0, \xi)$ and $\mG_{\rm{boe}}(\hD_1, \xi)$ cannot be expanded using the Mathematica package HypExp \cite{Huber:2005yg} since it only expands hypergeometric functions with parameters linear terms in $\e$. However, the same algorithm can still be used.\footnote{We explain this algorithm and write out the expansions for the boundary blocks in appendix \ref{App: Block Exp}.} 

Let us write the bootstrap equation in the following way
\begin{equation} \label{Bootstrap eq, eps 2}
\begin{aligned}
F_{\pm\pm}(\xi) &= G_b(\xi) + H_b(\xi) + \mco(\e^3) = G_i(\xi) + H_i(\xi) + \mco(\e^3) \ , \\
\end{aligned}
\end{equation}
where $H_b$ and $H_i$ contain the contributions from new operators
\begin{equation}
\begin{aligned}
G_b(\xi) &= 1 + \left( \e(\al + \bet) + \e^2\la a_0^{(2)} \right) \mG_{\rm{ope}}(\D_0; \xi) + \frac{\e\al}{2} \mG_{\rm{ope}}(\D_1; \xi) \ , \\
G_i(\xi) &= (1 + \e\bet)\xi^{\Delta_\phi}\mG_{\rm{boe}}(\hD_0; \xi) + \frac{\D_\ph^{(free)} + \e(\al - \bet)}{2}\xi^{\Delta_\phi}\mG_{\rm{boe}}(\hD_1; \xi) \ , \\
H_b(\xi) &= \e^2\sum_{n\geq 1}\la a_n^{(2)}\mG_{\rm{ope}}(2(n + 1); \xi) \ , \\
H_i(\xi) &= \e^2\sum_{m\geq 0}\m_m^{(2)}\xi\, \mG_{\rm{boe}}(m + 1; \xi) \ .
\end{aligned}
\end{equation}
We included the $\la a_0^{(2)}$-term in $G_b$ for a technical reason: the corresponding conformal block $\mG_{\rm{ope}}(2; \xi)$ has no branch cut along $\xi < - 1$. $G_b$ and $G_i$ have branch cuts along the entire negative axis $\xi < 0$ due to the non-integer powers of $\xi$, while the summand of $H_b$ has a branch cut along $\xi < -1$ and the summand of $H_i$ has a branch cut along $\xi \in (-1, 0)$. The branch cuts in $H_b$ and $H_i$ follow from the hypergeometric functions in the conformal blocks \eqref{Bootstrap eq} (see figure \ref{fig:analytic_structure_gi_gb}). We will consider the discontinuity \eqref{Disc bulk block} at $\xi < -1$, and assume that it commutes with the series' in $H_b$ and $H_i$ \eqref{Discs}. We can then use the orthogonality relation \eqref{Orth Rel} to project out the bulk OPE coefficients \eqref{Bulk coeff}
\begin{equation*} %\hspace{-30px}
\begin{aligned}
\la a_n^{(2)} &= \frac{\G_{n}\G_{n + 2}}{4\pi i(-1)^n\G_{2n + 1}} \int_{-1}^{+1}dy(1 - y)P_{n - 1}^{(1, 0)}(y) \left. \underset{{\xi < -1}}\disc[G_i(\xi) - G_b(\xi)] \frac{}{} \right|_{\xi = -\frac{2}{y + 1}} \\
&= \frac{(-1)^n(n!)^2}{(2n)!}\int_{-1}^{+1}dyP_{n - 1}^{(1, 0)}(y) \left\{ A_1 + (1 - y) \left[ A_2 + A_3\log\left(\frac{2}{1 + y}\right) + A_4 \log\left(\frac{1 - y}{1 + y}\right) \right] \right\} \ , \\
\end{aligned}
\end{equation*}
\begin{equation}
\begin{aligned}
A_1 &= -\frac{2\al\bet - (\al + \bet)\g_0^{(1)} + \al\g_1^{(1)} - \hg_{0}^{(2)} + \hg_1^{(2)}}{2} \ , \\
A_2 &= -\frac{\al(1 - 6\g_1^{(1)}) + 4\g_\ph^{(2)} - 2(\hg_{0}^{(2)} + \hg_1^{(2)})}{8} \ , \\
A_3 &= -\frac{\al\g^{(1)}_1}{4} \ , \\
A_4 &=  -\frac{\al(2\al - \g^{(1)}_1)}{4} \ .
\end{aligned}
\end{equation}
These integrals can be computed  to give %for each value of $n$ individually, and then the general form can be found % using the Mathematica command FindSequenceFunction \PDnote{Do we need to mention FindSequenceFunction?}
\begin{equation} \label{Bulk OPE coeff}
\begin{aligned}
\la a_1^{(2)} &= -\left( A_1 + A_2 + \frac{3}{2} A_3 + A_4 \right) \ , \\
\la a_{n \geq 2}^{(2)} &= \frac{2(n!)^2}{(2n)!} \left( \frac{(-1)^n A_1 }{n} - \frac{2 A_3 }{n^2 - 1} - \frac{2 A_4 }{n(n - (-1)^n)} \right) \ .
\end{aligned}
\end{equation}
We can now proceed to resum the bulk-channel using the procedure in appendix \ref{App: Bulk Resum}
\begin{equation*} \hspace{-10px}
\begin{aligned}
H_b %&= -\e^2 \left( A_1\frac{\xi}{\xi + 1}\log(\xi + 1) - 2A_2 \left( \frac{\xi}{\xi + 1} - \log(\xi + 1) \right) \right.\nn & \left.\, - 2A_3 \left( \frac{\xi}{\xi + 1} - \Li_2(-\xi) \right) + A_4 \log(\xi + 1)^2 \right) \,\nn
= -\e^2 \left( \frac{\xi}{\xi + 1} \bigg( A_1\log(\xi + 1) -2(A_2+A_3) \bigg)-2A_3 \Li_2(-\xi)+ 2A_2 \log(\xi + 1) + A_4 \log(\xi + 1)^2 \right)\,.%.\nn & \left.\, - 2A_3 \left( \frac{\xi}{\xi + 1}+ A_4 \log(\xi + 1)^2 - \Li_2(-\xi) \right) + A_4 \log(\xi + 1)^2 \right) \ .
\end{aligned}
\end{equation*}
As we can see, $H_b$ has a branch cut along $\xi < -1$ and none along $\xi \in (-1, 0)$, which coincides with our assumptions when we commuted the discontinuity with the series in $H_b$, see \eqref{Discs}. 

The $H_i$ we find from the bootstrap equation \eqref{Bootstrap eq, eps 2} contains terms with a branch cut along $\xi < -1$
\begin{equation}
\begin{aligned}
H_i \ni \al\frac{\g_1^{(1)} - 1}{2} \left( \log(\xi + 1) \left( \log\left(\frac{\xi}{\xi + 1}\right) + \frac{\log(\xi + 1)}{2} \right) + \Li_2(-\xi) \right) \ .
\end{aligned}
\end{equation}
These terms are not allowed since we assumed in \eqref{Discs} that it only has a branch cut along $\xi \in (-1, 0)$. If we demand these problematic terms to vanish we find
\begin{equation} \label{phi^4 theory}
\begin{aligned}
\al\g_1^{(1)} &= \al \ .
\end{aligned}
\end{equation}
That is, either $\al = 0$ or $\g_1^{(1)} = 1$. The latter is indeed the correct anomalous dimension of $\ph^j\pa^2\ph^j$ \cite{PhysRevD.7.2911}. We will assume \eqref{phi^4 theory} for generality, and remarkably this removes the dependence of $\g_1^{(1)}$ in the bootstrap equation. With this constraint at hand, we can project out the IOE coefficients using the orthogonality relation \eqref{Boundary Orth Rel}
% from $H_i$
\begin{equation}
\begin{aligned}
\m_m^{(2)} &= \underset{|w| = \tilde{\e}}{\oint}\frac{dw}{2\pi i}\frac{{}_2F_1(1 - m, -m, 2(1 - m), -w)}{w^{m + 1}} \left( \frac{B_1}{w + 1} + \frac{w B_2}{w + 1} + \rig \\
\eq \lef + \left( B_3 + \frac{B_4}{w + 1} \right) \log(w + 1) + B_5 \left[ \log(w + 1)^2 + 2\Li_2(-w) \right] \right) \ .
\end{aligned}
\end{equation}
\begin{equation}
\begin{aligned}
B_1 &= -\frac{2\g_\ph^{(2)} - \hg_{0}^{(2)} - \hg_{1}^{(2)} - 2\la a_0^{(2)}}{2} \ , \quad B_2 = -\frac{(2\al - \bet)(1 - 2\al) + 2\hg_1^{(2)}}{4} \ , \\
B_3 &= \g_\ph^{(2)} \ , \quad
B_4 = \frac{\al(1 + 2\bet) - (\al + \bet)\g_0^{(1)} - \hg_0^{(2)} + \hg_1^{(2)}}{2} \ , \quad B_5 = -\frac{\al(1 - 2\al)}{4} \,.
\end{aligned}
\end{equation}
These integrals can be performed which result in the following % for each n individually, and a general form can be found by using the Mathematica command FindSequenceFunction \PDnote{Do we need to mention FindSequenceFunction?}
\begin{equation} \label{BOE coeff}
\begin{aligned}
\m_0^{(2)} &= B_1 \ , \\
\m_1^{(2)} &= -B_1 + B_2 + B_3 + B_4 - 2B_5  \ , \\
\m_{m\geq 2}^{(2)} &= \frac{(m - 2)!}{2^{2m - 3}(3/2)_{m - 2}} \left( B_3 - (-1)^mB_4 - \frac{2(-1)^mB_5}{m(m - 1)} \right) \,,
\end{aligned}
\end{equation}
where $(a)_b = \frac{\G(a+b)}{\G(b)}$ is the Pochhammer symbol.
%Let us call even operators those with $m$ being even, and odd operators those with $m$ being odd. 
We can proceed to resum even and odd operators separately using the procedure explained in appendix \ref{App: Boundary Resum}. This yields two functions: $H_i^+$ and $H_i^-$ which corresponds to even and odd boundary operators respectively. Both of these satisfy the transformation property \eqref{Image symm} on their own
\begin{equation}
\begin{aligned}
H_i(\xi) = H_i^+(\xi) + H_i^-(\xi) \ , \quad \left. \frac{H_i^\pm}{\xi} \right|_{\xi \rightarrow -(\xi + 1)} = \pm \frac{H_i^\pm}{\xi} \ .
\end{aligned}
\end{equation}
This property is trivially satisfied for even operators ($H_i^+$) while it yields a constraint on the CFT data for odd operators ($H_i^-$)
\begin{equation}
\begin{aligned}
B_1 = 0 \ .
\end{aligned}
\end{equation}
This constraint together with \eqref{Bulk OPE coeff}, \eqref{phi^4 theory} and \eqref{BOE coeff} are the constraints we find from the bootstrap equation \eqref{Bootstrap eq}. They allow us to express all of the OPE coefficients in terms of the anomalous dimensions and the free parameters $\al$ and $\bet$.

\iffalse
\begin{equation}
\begin{aligned}
\la a_0^{(2)} &= \frac{2\g_\ph^{(2)} - \hg_0^{(2)} - \hg_1^{(2)}}{2} \ , \\
\la a_1^{(2)} &= \frac{2\g_\ph^{(2)} + 2\al(\al - \g_0^{(1)}) - 3\hg_0^{(2)} + \hg_1^{(2)}}{4} \ , \\
\la a_{n \geq 2}^{(2)} &= \frac{(n!)^2}{(2n)!} \left( \frac{\al}{n^2 - 1} - \frac{\al(1 - 2\al)}{n(n - (-1)^n)} - (-1)^n\frac{\al(1 - \g_0^{(1)}) - \hg_0^{(2)} + \hg_1^{(2)}}{n} \right) \ , \\
\end{aligned}
\end{equation}
\begin{equation}
\begin{aligned}
\m_0^{(2)} &= 0 \ , \\
\m_1^{(2)} &= \frac{2\g_\ph^{(2)} + \al(1 - \g_0^{(1)}) - \hg_0^{(2)}}{2} \ , \\
\m_{m \geq 2}^{(2)} &= \frac{(n!)^2}{4^{n - 1}(2n)!} \left( 2\g_\ph^{(2)} - (-1)^n(\al(1 - \g_0^{(1)}) - \hg_0^{(2)} + \hg_1^{(2)}) + (-1)^n\frac{\al(1 - 2\al)}{n(n - 1)} \right) \ .
\end{aligned}
\end{equation}
\fi

\subsection{Constraints from the equation of motion} \label{Sec: e.o.m.}

Let us now apply the e.o.m. to the BOE (or IOE), and study how we can constrain the CFT data from the previous subsection further. We will prove that only $\hp$ and $\pa_\perp\hp$ may appear in a free scalar theory of any unitary BCFT with a Lagrangian description. 
%In the process we will prove that only $\hp$ and $\pa_\perp\hp$ may appear in the free scalar theory of any unitary BCFT with a Lagrangian description. 
At $\mco(\e)$ we find the anomalous dimensions of these operators. We will not explore beyond $\mco(\e)$ in this section due to mixing that will occur at $\mco(\e^2)$. This is done in a similar way as in \cite{Giombi:2020rmc}.

The method we present is quite general, so we will not only do it for the ICFT we are considering, but also for a BCFT near four and three dimensions satisfying either Neumann or Dirichlet b.c.'s. See table \ref{Table: CFT data} for our final result on the anomalous dimensions. We were not able to constrain the anomalous dimensions in the $\ph^3$-theory near six dimensions in this way.

The BOE of $\ph$ is given by \cite{Billo:2016cpy}
\begin{equation} \label{BOE}
\begin{aligned}
\ph^i(x) &= \sum_{\hO}\sum_{m\geq 0}\frac{\m_\hD a_{\hD, m}}{|z|^{\D_\ph - \hD - 2m}}\pa_\para^{2m}\hO^i(x_\para) \ , \quad a_{\hD m} = \frac{(-1)^m}{4^mm!\left(\hD - \tfrac{d - 3}{2}\right)_m} \ .
\end{aligned}
\end{equation}
Here the BOE coefficients $\m^{\ph^i}{}_{\hO^j} \equiv \m_\hD\del^{ij}$ are to be treated as $O(N)$-matrices, which we have assumed to be proportional to the identity. The BOE yields the following two-point function
\begin{equation} \label{2pt fcn}
\begin{aligned}
\langle\ph^i(x)\hp^j(y)\rangle &= \del^{ij}\m_\hD\sum_{m\geq 0}\frac{a_{\hD m}}{|z|^{\D_\ph - \hD - 2m}}\pa_\para^{2m}\frac{A_d}{|s_\para|^{2\hD}} \ , \\ 
A_d &= \frac{1}{(d - 2)S_d} \ , \quad S_d = \frac{2\pi^{d/2}}{\G_{d/2}} \ , %\ , \quad s_\para^a \equiv x_\para^a - y_\para^a \ ,
\end{aligned}
\end{equation}
where %$\G_x$ is the shorthand notation of the Gamma-function $\G_x \equiv \G(x)$, 
$S_d$ is the area of a $(d - 1)$-dimensional sphere. We assumed the boundary fields live in an orthogonal basis
\begin{equation}
\begin{aligned}
\langle\hO_1^i(x_\para)\hO_2^j(y_\para)\rangle = \frac{A_d}{|s_\para|^{2\hD}}\del_{\hO_1^i\hO_2^j} \ .
\end{aligned}
\end{equation}
The e.o.m. is found by varying the action w.r.t. $\ph$, where we assume the bulk to have a sextic interaction near three dimensions and a quartic near four dimensions \cite{lawrie1984tricriticality, PhysRevD.7.2911} %and a cubic in six dimensions ($\sim\ph^3$) (In six dimensions there is neither a $O(N)$- nor a $\Z_2$-symmetry, in which case we set $N = 1$)
\begin{equation} \label{e.o.m.}
\begin{aligned}
\pa^2\ph^i = \la_n\ph^{a_n - 2}\ph^i \ , \quad \la_n = \frac{8\pi^2\e}{3N + 22}\de_{n, 3} + \frac{8\pi^2\e}{N + 8}\de_{n, 4} \ , \quad a_n = 6\de_{n, 3} + 4\de_{n, 4} \ .
\end{aligned}
\end{equation}
Here $n$ denotes the integer part of the spacetime dimension, $\la_n$ is the WF coupling constant at the RG fixed point and $\ph^{a_n - 2}$ is a composite operator. By dimensional analysis of the corresponding Lagrangian, $\D_\ph^{(free)}$ is given by \eqref{Free scaling dim}. If we let the derivatives on the LHS of the e.o.m. act on the correlator \eqref{2pt fcn}
\begin{equation} \label{LHS}
\begin{aligned}
\lhs &= \langle\pa^2\ph^i(x)\ph^j(y)\rangle = \pa_{x_\para}^2\langle\ph^i(x)\ph^j(y)\rangle + \pa_{z}^2\langle\ph^i(x)\ph^j(y)\rangle \\
&= \del^{ij}\m_\hD\sum_{m\geq 0}\frac{a_{\hD, m - 1} + c_{\hD, m}a_{\hD, m}}{|z|^{\D_\ph - \hD - 2m + 2}}\pa_\para^{2m}\frac{1}{|s_\para|^{2\hD}} \ , \\
c_{\hD, m} &= (\D_\ph - \hD - 2m)(\D_\ph - \hD - 2m + 1) \ .
\end{aligned}
\end{equation}
The summand can be simplified to
\begin{equation} \hspace{-6px}\label{LHS simplified}
\begin{aligned}
&a_{\hD, m - 1} + c_{\hD, m}a_{\hD, m} %= \\
%\eq 
= \frac{(\D_\ph - \hD)^2 + (\D_\ph - \hD) + 2m \left[ d - 2(\D_\ph + 1) \right]}{(-4)^mm!}\frac{\G_{\hD - (d - 3)/2}}{\G_{m + \hD - (d - 3)/2}} \ .
\end{aligned}
\end{equation}
We can choose a renormalization scheme such that the one-point function of $\langle\ph^2\rangle$ is the finite part of the coincident limit of $\langle\ph(x)\ph(y)\rangle$. We assume it to be given by
\begin{equation}
\begin{aligned}
\langle \ph^2(x)\rangle &= \frac{\kap_n^\pm}{|z|^{2\D_\ph}} + \mco(\la_n) \ .
%= \pm\frac{\del^{ij}}{8\pi|z|}\del_{n3} \pm \frac{\del^{ij}}{16\pi^2z^2}\del_{n4} + \mco(\la_n, \e) \ .
%\fin\lim\limits_{y\rightarrow x}\langle\ph^i(x)\ph^j(y)\rangle =
\end{aligned}
\end{equation}
The coefficient $\kap_{n}^\pm$ can be found in table \ref{Table: Anom Dim} for different models. The RHS of the e.o.m. \eqref{e.o.m.} acting on the correlator \eqref{2pt fcn} is found using Wick's theorem
\begin{equation}
\begin{aligned}
\rhs &= \langle\la_n\ph^{a_n - 2}\ph^i(x)\ph^j(y)\rangle = \del^{ij}\la_n\si_n\langle\ph^2(x)\rangle^{a_n - 2}\langle\ph(x)\ph(y)\rangle \\
&= \del^{ij}\m_\hD\sum_{m\geq 0} \left( a_{\hD, m}\frac{\la_n\si_n(\kap_{n}^\pm)^{a_n - 2}}{|z|^{\D_\ph - \hD - 2m + 2}} \pa_\para^{2m}\frac{A_d}{|s_\para|^{2\hD}} + \mco(\la_n^2, \la_n\e) \right) \ , \\
\si_n &= (N + 4)(N + 2)\de_{n, 3} + (N + 2)\de_{n, 4} \ .
\end{aligned}
\end{equation}
Here $\si_n$ is the symmetry factor from Wick's theorem. We can compare this to the LHS \eqref{LHS} to find an equation for the scaling dimensions that holds for all $m \in\Z_{\geq 0}$ up to order $\e$
\begin{equation} \label{BOE e.o.m. Eq}
\begin{aligned}
&a_{\hD, m - 1} + c_{\hD, m}a_{\hD, m} = a_{\hD, m}\la_n\si_n(\kap_{n}^\pm)^{a_n - 2} \ .
\end{aligned}
\end{equation}
Let us first solve this equation for the free theory when the LHS is zero on its own. Then either the $\G$-functions are zero or the numerator is zero in equation \eqref{LHS simplified}. The $\G$-functions are zero when $\hD = \D_\ph^{(free)} - (m + k) \ , m + k \geq 1 \ , m , k \in \Z_{\geq 0}$. However, these solutions are not valid if the CFT on the boundary is unitary. 

The numerator on the other hand is zero when
\begin{equation}
\begin{aligned}
\hD^{(free)}_+ = \D_\ph^{(free)} \ , \quad\text{or ,}\quad \hD^{(free)}_- = \D_\ph^{(free)} + 1 \ .
\end{aligned}
\end{equation}
These are the scaling dimensions of $\hp$ and $\pa_\perp\hp$. In a BCFT they correspond to Neumann or Dirichlet b.c.'s respectively. This proves that these are the only operators that can appear in the BOE of $\ph$ in the free theory. Note that this result is independent of the spacetime dimension and also applies to ICFTs. This result is not new in itself, and has been found prior to this work in \cite{Gliozzi:2015qsa, Giombi:2020rmc}.

In the interacting theory, we expand the scaling dimensions in $\e$, using as input from the bulk theory that the fundamental scalar in the bulk has no anomalous dimension at this order
\begin{equation} %\hspace{-10px}
\begin{aligned}
\D_\ph &= \D_\ph^{(free)} + \mco(\e^2) \ , \\ 
\hD_\pm &= \hD^{(free)}_\pm + \e(\g_{n}^\pm)^{(1)} + \mco(\e^2) \ .
\end{aligned}
\end{equation}
The e.o.m. \eqref{BOE e.o.m. Eq} has one solution that holds for any $m$
\begin{equation}
\begin{aligned}
(\g_{n}^\pm)^{(1)} &= \mp\frac{\la_n\si_n(\kap_{n}^\pm)^{a_n - 2}}{\e} \ .
\end{aligned}
\end{equation}
We list the anomalous dimensions for different models in table \ref{Table: Anom Dim}. The BCFT results near three and four dimensions match with the existing literature \cite{PhysRevLett.45.1581, REEVE1981237, Diehl1981FieldtheoreticalAT, 1983ZPhyB..51..361S, Diehi_1987}.

\begin{table}[]
	\centering
	\begin{tabular}{|l|l|l|l|}
		\hline
		\textbf{Type of defect:} 	& $n$ & $\kap_n^{\pm}$                                        		& $(\g_{n}^\pm)^{(1)}$					\\ \hline
		Boundary 					& $3$ & $\left(\pm\tfrac{A_3}{2}\right)^2 = \tfrac{1}{64\pi^2}$ 	& $\mp\frac{(N + 4)(N + 2)}{8(3N + 22)}$	\\ \hline
		Boundary 					& $4$ & $\pm\frac{A_4}{2} = \pm\frac{1}{16\pi^2}$             		& 
		$-\frac{N + 2}{2(N + 8)}$				\\ \hline
		Interface 					& $4$ & $0$															& 
		$0$										\\ \hline
	\end{tabular}
	\caption{The table shows the one-point functions of $\ph^2$ and anomalous dimensions in different models: $n$ denotes the integer part of the spacetime dimension, and the subscript "$\pm$" only applies to BCFTs where "$+$" means Neumann b.c., and "$-$" means Dirichlet.} \label{Table: Anom Dim} \label{Table: CFT data}
\end{table}

Of particular interest for us, the anomalous dimensions of $\hp$ and $\pa_\perp\hp$ in the ICFT that was considered in the previous section are both zero.\footnote{One can also see this by letting the e.o.m. act on the resummed correlator at order $\e$.} Thus from \eqref{Anom Dim} we can deduce that
\begin{equation}
\begin{aligned}
\al &= 0 \ .
\end{aligned}
\end{equation}
This is the last constraint we find on the CFT data in the ICFT we are considering. 

Let us remind the reader that we consider another interface than those studied in \cite{BM77, Sym81, BE81, EB82, DDE83}, where the anomalous dimensions of $\hp$ and $\pa_\perp\hp$ are not zero.

\subsection{CFT data at order $\e^2$}\label{CFTdatasummary}

In this subsection we summarize the CFT data related to the interface
\iffalse
\begin{equation}
\begin{aligned}
\D_\ph &= \D_\ph^{(free)} + \e^2\g_\ph^{(2)} + \mco(\e^3) \ , \\
\end{aligned}
\end{equation}

\begin{equation}
\begin{aligned}
% \D_0 \equiv \D_{\ph^2} &= 2\D_\ph^{(free)} + \e\g_0^{(1)} + \mco(\e^2) \ , \\
%\D_1 \equiv \D_{\ph\pa^2\ph} &= 4 + \mco(\e^2) \ , \\
\hD_0 \equiv \D_\hp &= \D_\ph^{(free)} + \e^2\hg_0^{(2)} + \mco(\e^3) \ , \\
\hD_1 \equiv \D_{\pa\hp} &= \D_\ph^{(free)} + 1 + \e^2\hg_1^{(2)} + \mco(\e^3) \ , \\
\end{aligned}
\end{equation}
\fi

\iffalse
\begin{equation} %\label{Bulk OPE coeffs}
\begin{aligned}
\la a_0 &= \e\beta+\e^2\frac{2\g_\ph^{(2)} - \hg_0^{(2)} - \hg_1^{(2)}}{2} + \mco(\e^3) \ , \\
\la a_1 &= -\e^2\frac{2\bet \g_0^{(1)} - 2\g_\ph^{(2)} + 3\hg_0^{(2)} - \hg_1^{(2)}}{4} + \mco(\e^3) \ , \\
\la a_{n \geq 2} &= \e^2\frac{(-1)^n(n!)^2}{(2n)!}\frac{\bet \g_0^{(1)} + \hg_0^{(2)} - \hg_1^{(2)}}{n} + \mco(\e^3) \ , \\
\end{aligned}
\end{equation}

\begin{equation} %\label{BOE coeffs}
\begin{aligned}
\m_0 &= 1 + \mco(\e^3) \ , \\
\m_1 &= \frac{\D_\ph^{(free)}}{2} + \e^2\frac{\bet \left(1 - 2\g_0^{(1)} \right) + 4\g_\ph^{(2)} - 2\hg_0^{(2)}}{4} + \mco(\e^3) \ , \\
\m_{m \geq 2} &= \e^2\frac{(m - 2)!}{(3/2)_{m - 2}} \frac{2\g_\ph^{(2)} + (-1)^m \left( \bet \g_0^{(1)} + \hg_0^{(2)} - \hg_1^{(2)} \right) }{4^{m - 1}} + \mco(\e^3)\ .
\end{aligned}
\end{equation}
\fi

\begin{equation} \label{Bulk OPE coeffs}
\begin{aligned}
\la a_0 &= \e\beta+\e^2\frac{2\g_\ph^{(2)} - \hg_0^{(2)} - \hg_1^{(2)}}{2} + \mco(\e^3) \ , \\
\la a_1 &= -\e^2\frac{2\bet \g_0^{(1)} - 2\g_\ph^{(2)} + 3\hg_0^{(2)} - \hg_1^{(2)}}{4} + \mco(\e^3) \ , \\
\la a_{n \geq 2} &= \e^2\frac{(-1)^n(n!)^2}{(2n)!}\frac{\bet \g_0^{(1)} + \hg_0^{(2)} - \hg_1^{(2)}}{n} + \mco(\e^3) \ ,
\end{aligned}
\end{equation}
\begin{equation} \label{Bulk OPE coeffs 2}
\begin{aligned}
\m_0^2 &= 1 + \mco(\e^3) \ , \\
\m_1^2 &= \frac{\D_\ph^{(free)}}{2} + \e^2\frac{\bet \left(1 - 2\g_0^{(1)} \right) + 4\g_\ph^{(2)} - 2\hg_0^{(2)}}{4} + \mco(\e^3) \ , \\
\m_{m \geq 2}^2 &= \e^2\frac{(m - 2)!}{(3/2)_{m - 2}} \frac{2\g_\ph^{(2)} + (-1)^m \left( \bet \g_0^{(1)} + \hg_0^{(2)} - \hg_1^{(2)} \right) }{4^{m - 1}} + \mco(\e^3)\ .
\end{aligned}
\end{equation}
The full correlator expressed in the anomalous dimensions is given by
\begin{equation*}
\begin{aligned}
F_{\pm\pm}(\xi) &= 1 + \e\bet\frac{\xi}{\xi + 1} - \frac{\e^2}{2} \left( \bet ( 1 - \g_0^{(1)} ) \frac{\xi}{\xi + 1}\log(\xi) + \rig \\
\eq \lef - \left( (2\g_\ph^{(2)} - \hg_0^{(2)} - \hg_1^{(2)}) + ( \bet ( 1 - \g_0^{(1)} ) - \hg_0^{(2)} + \hg_1^{(2)} ) \frac{\xi}{\xi + 1} \right) \log(\xi + 1) \right) + \mco(\e^3) \ .
\end{aligned}
\end{equation*}
We can resum the correlator in $\e$ to find
\begin{equation} \label{Full Corr}
\begin{aligned}
\langle\ph_\pm^i(x)\ph_\pm^j(y)\rangle &= \de^{ij} \left( \frac{(\xi + 1)^{\g_\ph - (\hg_0 + \hg_1)/2}}{|x - y|^{2\D_\ph}} + %\rig \\
%\eq\lef 
 \e\bet\frac{(\xi + 1)^{(\hg_1 - \hg_0)/{2\bet}}}{|\tilde{x} - y|^{2\D_\ph}} \left( \frac{\xi}{\xi + 1} \right)^{\g_0/2 - \g_\ph} \right) + \mco(\e^3) \ .
% \langle\ph_\pm^i(x)\ph_\pm^j(y)\rangle &= \de^{ij} \left( \frac{(\xi + 1)^{\g_\ph - (\hg_0 + \hg_1)/2}}{|x - y|^{2\D_\ph}} - \frac{(\xi + 1)^{\hg_0/2} - (\xi + 1)^{\hg_1/2}}{|\tilde{x} - y|^{2\D_\ph}} \right) + \mco(\e^3) \ .
\end{aligned}
\end{equation}
Here $\D_\ph$ is the full scaling dimension of $\ph$ (with the anomalous dimensions included), $\g_\ph, \g_0, \hg_0$ as well as $\hg_1$ are the full anomalous dimensions (to all orders of $\e$), and $\tilde{x} = \left.x\right|_{z \rightarrow -z}$ is the image point of $x$. Note that this correlator does not satisfy the image symmetry \\
$(z, z')\rightarrow(-z, -z')$. To our knowledge the anomalous dimensions of the interface fields are not known. The anomalous dimensions of $\ph$ and $\ph^2$ are known \cite{PhysRevD.7.2911} and can be used as input from the bulk theory
\begin{equation}
\begin{aligned}
\g_0 &= \frac{N + 2}{N + 8}\e + \mco(\e^2) \ , \quad &\g_\ph &= \frac{N + 2}{4(N + 8)^2}\e^2 + \mco(\e^3) \ .
\end{aligned}
\end{equation}

\newpage

\subsection{The renormalization group domain wall}

Let us now specialize to the case when one side of the interface is free, say the side with $\ph_-$ at $z < 0$. This model was studied in \cite{Gliozzi:2015qsa}. Since the interface fields are limits of the free fields in the bulk at $z < 0$ (as seen from the b.c.'s \eqref{Boundary conditions}), they are all protected. This means that we can retrieve the full $\ph_+-\ph_+$ correlator as well as the OPE coefficients from \eqref{Bulk OPE coeffs}, \eqref{Bulk OPE coeffs 2} and \eqref{Full Corr} by setting the interface anomalous dimensions to zero\footnote{We checked this result using the same bootstrap procedure as in section \ref{Sec: ICFT Bootstrap}.}
\begin{equation}
\begin{aligned}
\hg_{m} = 0 \ \forall\ m\in\Z_{\geq 0} \ .
\end{aligned}
\end{equation}
This yields the correlator
\begin{equation}
\begin{aligned}
\langle\ph_+^i(x)\ph_+^j(y)\rangle &= \de^{ij} \left( \frac{(\xi + 1)^{\g_\ph}}{|x - y|^{2\D_\ph}} + \frac{\e\bet}{|\tilde{x} - y|^{2\D_\ph}} \left( \frac{\xi}{\xi + 1} \right)^{\g_0/2 - \g_\ph} \right) + \mco(\e^3) \ .
\end{aligned}
\end{equation}
\iffalse
\begin{equation}
\begin{aligned}
\la a_0 &= \e\bet + \g_\ph + \mco(\e^3) \ , \quad &\la a_1 &= -\frac{\e\bet\g_0 - \g_\ph}{2} + \mco(\e^3) \ , \\
\m_{0}^2 &= 1 + \mco(\e^3) \ , \quad &\m_1^2 &= \frac{\D_\ph^{(free)}}{2} + \g_\ph + \frac{\e\bet(\e - 2\g_0)}{4} + \mco(\e^3) \ , \\
\la a_n &= \g_\ph \left( \de_{n, 0} + \frac{\de_{n, 1}}{2} \right) + \mco(\e^3) \ , \quad &\m_{\geq 2}^2 &= \frac{\G_{n - 1}\g_\ph}{2^{2n - 3}(3/2)_{n - 2}} + \mco(\e^3) \ .
\end{aligned}
\end{equation}
\fi
We find $\langle \ph^i_-\ph^j_- \rangle$ from $\langle \ph^i_+\ph^j_+ \rangle$ by setting the bulk anomalous dimensions to zero
\begin{equation}
\begin{aligned}
\g_\ph &= 0 \quad\Rightarrow\quad \langle \ph^i_-\ph^j_- \rangle = \de^{ij}\left( \frac{1}{|x - y|^{2\D_\ph^{(free)}}} + \frac{\e\bet}{|\tilde{x} - y|^{2\D_\ph^{(free)}}} \right) + \mco(\e^3) \ .
\end{aligned}
\end{equation}
Note that even though both the bulk and the boundary operators are protected, this correlator still differs from that in a homogeneous CFT.

%\newpage

%%%%%%%%%%%%%%%%%%%%%%%%%%%%%%%%%%%%%%%%%%%%%%%%%%%%%%%%%%%%%%%%%%%%%%%%%%%%%%%

\section{A boundary CFT in $6-\e$ dimensions} \label{6d BCFT}

In this section we study the $\f^3$-theory in $d = 6 - \e$ dimensions close to a boundary. The bulk theory is called the Landau-Ginzburg model and it describes the distribution of "Yang-Lee zeros" (the zeros of the partition function) on the imaginary magnetic field axis in ferromagnets above the critical temperature \cite{Fis78}. The model is not unitary so the OPE coefficients do not need to be real-valued. 

This model was considered in the presence of a boundary \cite{JL95}, and a generalization of it with a tensorial coupling constant and $N$-vector fields was studied in \cite{Car80, DL89}. It describes the continuum limit of the $(N + 1)$-state Potts model, where $N \rightarrow 0$ corresponds to percolation, and $N = 1$ describes an extended Ising model. The case of $N = 1$ belongs to another universality class than $N \geq 2$, and thus our result may differ from those in \cite{Car80, DL89}.

%The $\f^3$ theory in $d=6-\e$ dimensions describes the distribution of Yang-Lee zeros on the imaginary magnetic field axis in ferromagnets above the critical temperature. This has been studied in \cite{Fis78, JL95}. A generalization of this model with a tensorial coupling constant and $N$-vector fields was studied in \cite{Car80, DL89}. It describes the continuum limit of the $(N + 1)$-state Potts model, where $N \rightarrow 0$ corresponds to percolation, and $N = 1$ describes an extended Ising model. The case of $N = 1$ belongs to another universality class than $N \geq 2$, and thus our result may differ from theirs.  

We consider the two-point function \eqref{2pt} of the fundamental scalar $\phi$ of dimension $\D_\f$, and bootstrap this correlator upto $\mco(\e)$. This two-point function in the free theory reads
\begin{align}\label{phi3corr}
F^{\pm}(\xi)= 1 \pm \left( \frac{\xi}{\xi + 1} \right)^{\D_\f^{(free)}} \ , \quad \D^{{(free)}}_\f = \frac{d - 2}{2} \ .
\end{align}
Here "$+$" and "$-$" represent Neumann and Dirichlet b.c.'s respectively
\begin{equation} \label{Neumann}
\begin{aligned}
\pa_\perp\hp = 0 \ ,
\end{aligned}
\end{equation}
%Here "$-$" represents Dirichlet BCs
\begin{equation} \label{Dirichlet}
\begin{aligned}
\hp = 0 \ .
\end{aligned}
\end{equation}
The $\ph^2$-operator is a descendant of $\ph$ in the interacting theory, which can be seen from the e.o.m. $\pa^2 \f \sim \f^2$. This means that the operator $\f$ appears in the bulk OPE of $\f \times \f$ instead of $\ph^2$ (this is a consequence of breaking $\Z_2$-symmetry). The bulk conformal block corresponding to the exchange of $\ph$ contains a simple pole in $\e$. This singularity is compensated by including the anomalous dimension of $\ph$ already in the free theory. The decomposition is%\footnote{One could consider a $\ph^2$-exchange in the bulk-channel in the free theory. Although such composition does not hold at higher orders in $\e$.}
\begin{equation}
\begin{aligned}
\la a_\1 &= 1 \ , \quad &\la a_\D^\pm &= \mp\e\g_\ph^{(1)}\de_{\D, \D_\ph} \ , \\ 
\m_\1^2 &= 0 \ , \quad &(\m^\pm_\hD)^2 &= (1 \pm 1)\del_{\hD, \D_\ph^{(free)}} + (1 \mp 1)\frac{\D_\ph^{(free)}}{2}\del_{\hD, \D_\ph^{(free)} + 1} \ ,
\end{aligned}
\end{equation}
where $\g_\ph^{(1)}$ is the anomalous dimension of $\ph$ at order $\e$. This block decomposition indicates that only the bulk identity and  $\f$ contribute in the bulk-channel whereas only  $\hat{\f}$ or $\partial_{\perp} \hat{\f}$  contributes in the boundary-channel with Neumann or Dirichlet b.c. respectively. This agrees with the results from section \ref{Sec: e.o.m.} even though this theory is non-unitary.

We will now use the bootstrap equation  \eqref{Bootstrap eq} to constrain the CFT data at higher orders in $\e$. The CFT data admits an expansion in $\sqrt{\e}$. One can check (either by expanding in $\xi$, or by using the method presented in this paper), that all of the CFT data in the bootstrap equation is zero at order $\sqrt{\e}$. We will thus expand the CFT data as
\begin{equation}
\begin{aligned}
\D_\ph &= \D_\ph^{(free)} + \e\g_\ph^{(1)} + \mco(\e^{3/2}) \ , \\
\D_{n\geq 1} &= 2 \left( \D_\ph^{(free)} + n \right) + \e\g_n^{(1)} + \mco(\e^{3/2}) \ , \\
\hD_{m\geq 2} \equiv \D_{\pa_\perp^m\hp} &= \D_\ph^{(free)} + m + \e\hg_m^{(1)} + \mco(\e^{3/2}) \ , \\
\end{aligned}
\end{equation}

\begin{equation}
\begin{aligned}
\la a_\1 &= 1 \ , \\
\la a_\ph^\pm &= \mp\e\g_\ph\de_{\D, \D_\ph} + \e^2(\la a_\ph^{(2)})^\pm + \mco(\e^{5/2}) \ , \\
\la a_{\D_n}^\pm &= \e(\la a_n^{(1)})^\pm + \mco(\e^{3/2}) \ , \\
\end{aligned}
\end{equation}

\begin{equation}
\begin{aligned}
\m_\1^2 &= 0 \ , \\ 
\m^2_{\hD_m} &= (1 \pm 1)\del_{m, 0} + (1 \mp 1)\frac{\D_\ph^{(free)}}{2}\del_{m, 1} + \e(\m_m^{(1)})^\pm + \mco(\e^{3/2}) \ .
\end{aligned}
\end{equation}
Let us write the bootstrap equation at order $\e$ as
\begin{equation} \label{beqe}
\begin{aligned}
G_b^\pm(\xi) + H_b^\pm(\xi) + \mco(\e^{3/2}) = G_i^\pm(\xi) + H_i^\pm(\xi) + \mco(\e^{3/2}) \ ,
\end{aligned}
\end{equation}
where $H_b$ and $H_i$ correspond to new operators
\begin{equation}
\begin{aligned}
G_b^\pm(\xi) &= \left( \mp\e\g_\ph^{(1)}\de_{\D, \D_\ph} + \e^2(\la a_\ph^{(2)})^\pm  \right) \mG_{\rm{ope}}(\D_\ph; \xi) \ , \\
G_i^\pm(\xi) &= (1 \pm 1)\mG_{\rm{boe}}(\hD_0; \xi) + (1 \mp 1)\frac{\D_\ph^{(free)}}{2}\mG_{\rm{boe}}(\hD_1; \xi) \ , \\
H_b^\pm(\xi) &= \e \sum_{n \geq 1} (\la a_n^{(1)})^\pm \mG_{\rm{ope}}(2(n + 1); \xi) \ , \\
H_i^\pm(\xi) &= \e \sum_{m \geq 0} (\m_m^{(1)})^\pm \mG_{\rm{boe}}(m + 1; \xi) \ .
\end{aligned}
\end{equation}
We included the bulk $\ph$-exchange in $G_b$ since its conformal block does not have a branch cut along $\xi < -1$. 

Now take the discontinuity \eqref{Disc bulk block} around $\xi < -1$, and assume that it commutes with the series' in $H_b$ and $H_i$ \eqref{Discs}. We proceed to project out the bulk OPE coefficients \eqref{Bulk coeff} using the orthogonality relation \eqref{Orth Rel} of the Jacobi polynomials
\begin{equation}
\begin{aligned}
(\la a_n^{(1)})^\pm &= \frac{(-1)^n(n - 1)!(n + 1)!}{(2n)!}\int_{-1}^{+1}dyP_{n - 1}^{(2, 0)}(y) \left( A_1^\pm + A_2^\pm (1 - y)^2 \right) \ , \\
A_1^\pm &= \frac{\pm (1 - 2\g_\ph^{(1)}) + 2(1 \pm 1)\hg_0^{(1)} - 2(1 \mp 1)\hg_1^{(1)}}{8} \ , \\
A_2^\pm &= -\frac{2\g_\ph^{(1)} - (1 \pm 1)\hg_0^{(1)} - (1 \mp 1)\hg_1^{(1)}}{16} \ .
\end{aligned}
\end{equation}
Performing the integral over $y$ yields
\begin{equation} \label{bulk6ddata 1}
\begin{aligned}
(\la a_1^{(1)})^\pm &= -2A_1^\pm - \frac{8}{3}A_2^\pm \ , \quad &(\la a_{n \geq 2}^{(1)})^\pm &= 2\frac{(-1)^n(n - 1)!(n + 1)!}{(2n)!}A_1^\pm \ .
\end{aligned}
\end{equation}
The bulk-channel is resummed using the integral representation \eqref{Int Repr}
\begin{equation*}
\begin{aligned}
H_b^\pm &= -\e \left( \frac{\xi}{\xi + 1} \left( 6A_1^\pm + \left( 5A_1^\pm + \frac{8}{3}A_2^\pm \right) \frac{\xi}{\xi + 1} \right) - \left( 6A_1^\pm - 2A_2^\pm \left( \frac{\xi}{\xi + 1} \right)^2 \right) \log(\xi + 1) \right) \ .
\end{aligned}
\end{equation*}
With this we can find the full correlator. If we impose Neumann or Dirichlet b.c.'s we also get rid of unwanted branch cuts along $\xi \in (-1, 0)$ in $H_i^\pm$ from the bootstrap equation \eqref{beqe}\footnote{For Neumann b.c.'s, there is a simple pole in $z$ in the boundary limit of $\pa_\perp\ph$, which corresponds to the expected exchange of $\hp$ in the BOE. This term do not need to be zero. Only the $z^0$-term corresponding to the exchange of $\pa_\perp\hp$ should be set to zero.}
\begin{equation} \label{bulk6ddata 2}
\begin{aligned}
(\la a_\ph^{(2)})^+ &= -\frac{17}{32}\g_\ph^{(1)}(1 + 2\g_\ph^{(1)}) \ , \quad &\hg_0^{(1)} &= -\frac{3}{16} + \frac{5}{8}\g_\ph^{(1)} \ , \\
(\la a_\ph^{(2)})^- &= -\frac{\g_\ph^{(1)}(1 + 2\g_\ph^{(1)})}{16} \ , \quad &\hg_1^{(1)} &= -\frac{3}{2} + \frac{\g_\ph^{(1)}}{4} \ .
\end{aligned}
\end{equation}
We project out the BOE coefficients from $H_i^\pm$ using the orthogonality relation \eqref{Boundary Orth Rel}
\begin{equation}
\begin{aligned}
(\m_m^{(1)})^\pm &= \underset{|w| = \tilde{\e}}{\oint}\frac{dw}{2\pi i}\frac{{}_2F_1(-1 - m, 1 - m, 2(1 - m), -w)}{w^{m + 1}} \left( B_1^\pm + B_2^\pm\frac{w}{w + 1} + \rig \\
\eq \lef + B_3^\pm \left( \frac{w}{w + 1} \right)^2 + \left( B_4^\pm + B_5^\pm\frac{w}{w + 1} + B_6^\pm \left( \frac{w}{w + 1} \right)^2 \right) \log(w + 1) \right) \ .
\end{aligned}
\end{equation}
\begin{equation}
\begin{aligned}
B_1^+ &= -\frac{1}{4} - \frac{\g_\ph^{(1)}}{2} \ , \quad &B_2^+ &= \frac{5}{16} - \frac{3}{8}\g_\ph^{(1)} \ , \quad &B_3^+ &= -\frac{1}{4} + \frac{3}{2}\g_\ph^{(1)} \ , \\ 
B_4^+ &= -\frac{1}{16} + \frac{7}{8}\g_\ph^{(1)} \ , \quad &B_5^+ &= \frac{1}{8} + \frac{\g_\ph^{(1)}}{4} \ , \quad &B_6^+ &= -\frac{1}{16} - \frac{\g_\ph^{(1)}}{8} \ , \\
\end{aligned}
\end{equation}
\begin{equation}
\begin{aligned}
B_1^- &= 0 \ , \quad &B_2^- &= 0 \ , \quad &B_3^- &= \frac{3}{16} - \frac{\g_\ph^{(1)}}{8} \ , \\ 
B_4^- &= -\frac{1}{8} + \frac{3}{4}\g_\ph^{(1)} \ , \quad &B_5^- &= \frac{1}{4} + \frac{\g_\ph^{(1)}}{2} \ , \quad &B_6^- &= -\frac{1}{8} - \frac{\g_\ph^{(1)}}{4} \ , \\
\end{aligned}
\end{equation}
Calculating the residues gives us the BOE coefficients
%\begin{equation}
\begin{equation}
\begin{aligned}
(\m_0^{(1)})^\pm &= B_1^\pm \, ,\\
(\m_1^{(1)})^- &= B_2^- + B_4^- \, ,\\ 
(\m_2^{(1)})^\pm &= \frac{B_2^\pm}{2} + B_3^\pm + B_4^\pm + B_5^\pm \, ,
\end{aligned}
\end{equation}
%\end{equation}
\begin{equation*} \hspace{-10px}
\begin{aligned}
(\m_{m \geq 3}^{(1)})^\pm &= \frac{\sqrt{\pi}\G_{m - 2}}{\G_{m - 1/2}}\frac{(m - 2)(m + 1)B_4^\pm + (m(m - 1) - 1 - (-1)^m)B_5^\pm + m(m - 1)(1 - (-1)^m)B_6^\pm}{4^{m - 1}} \ .
\end{aligned}
\end{equation*}
We have to be a bit careful as $\pa_\perp\hp$ (corresponding to $\m_1^+$) does not appear for Neumann b.c.'s (even though its coefficient seems to be non-zero from the calculations).\footnote{Physically the normal derivative is forbidden due to the Neumann b.c. \eqref{Neumann}.} One can see that this block should not appear by resumming the boundary-channel,\footnote{We resum even and odd parts separately using the procedure in appendix \ref{App: Boundary Resum}.} and comparing the result with the $H_i^+$ found from the bootstrap equation \eqref{beqe}. While resumming the boundary-channel we find that the property \eqref{Image symm} is trivially satisfied for both even and odd operators.

The BOE coefficients above, together with the bulk OPE coefficients and boundary anomalous dimensions at \eqref{bulk6ddata 1} and \eqref{bulk6ddata 2} are the constraints on the CFT data found from the bootstrap equation. They are all expressed in terms of one parameter: the anomalous dimension of the external field $\ph$. We were not able to implement the e.o.m. in such a way that we could find this parameter. Hence we will use it as an input \cite{priest1976critical, amit1976renormalization}\footnote{Note that these references consider $d = 6 - 2\e$ while we consider $d = 6 - \e$.}
%\cite{Gracey:2015tta} (CORRECT CITATION??)
\begin{equation}
\begin{aligned}
\g_\ph^{(1)} &= -\frac{1}{18} \quad\Rightarrow\quad (\la a_1^{(1)})^\pm = 0 \ .
\end{aligned}
\end{equation}
The non-trivial CFT data for Neumann b.c.'s is given by
\begin{equation}
\begin{aligned}
\hg_0 &= -\frac{2}{9}\e + \mco(\e^{3/2}) \ , \\
\la a_\ph^+ &= \frac{\e}{18} + \frac{17}{648}\e^2 + \mco(\e^{5/2}) \ , \\
\la a_{n\geq 2}^+ &= \frac{(-1)^n(n - 1)!(n + 1)!}{18(2n)!}\e + \mco(\e^{3/2}) \ ,
\end{aligned}
\end{equation}
\begin{equation}
\begin{aligned}
(\m_0^+)^2 &= 2 - \frac{2}{9}\e + \mco(\e^{3/2}) \ , \\
(\m_2^+)^2 &= -\frac{\e}{6} + \mco(\e^{3/2}) \ ,\\
(\m_{m\geq 3}^+)^2 &= -\frac{\sqrt{\pi}(m + 1)\G_{m - 1}(1 - (-1)^m)}{2^{2m - 1}\G_{m - 1/2}}\frac{\e}{9} + \mco(\e^{3/2}) \ ,
\end{aligned}
\end{equation}
where $\hg_0$ agrees with the result from \cite{JL95}, and the non-trivial CFT data for Dirichlet b.c.'s is given by
\begin{equation}
\begin{aligned}
\hg_1 &= -\frac{7}{18}\e + \mco(\e^{3/2}) \ , \\ 
\la a_\ph^- &= -\frac{\e}{18} + \frac{\e^2}{324} + \mco(\e^{5/2}) \ , \\
\la a_{n\geq 2}^- &= \frac{(-1)^n(n - 1)!(n + 1)!}{9(2n)!}\e + \mco(\e^{3/2}) \ ,
\end{aligned}
\end{equation}
\begin{equation}
\begin{aligned}
(\m_1^-)^2 &= \D_\ph^{(free)} - \frac{\e}{6} + \mco(\e^{3/2}) \ , \\
(\m_2^-)^2 &= \frac{\e}{4} + \mco(\e^{3/2}) \ , \\
(\m_{m\geq 3}^-)^2 &= -\frac{\sqrt{\pi}(m + 1)\G_{m - 1}(1 - 2(-1)^m)}{2^{2m - 1}\G_{m - 1/2}}\frac{\e}{9} + \mco(\e^{3/2}) \ .
\end{aligned}
\end{equation}
%Due to non-unitary, not all BOE coefficients are real. 
The full correlators reads
\begin{equation}
\begin{aligned}
F^+(\xi) &= 1 + \left( \frac{\xi}{\xi + 1} \right)^2 - \frac{\e}{3} \left( \frac{1}{3} \left( 1 + \frac{\xi}{\xi + 1} \right) \frac{\xi}{\xi + 1} + \frac{5}{6} \left( \frac{\xi}{\xi + 1} \right)^2 \log(\xi) + \rig \\
\eq \lef - \left( \frac{1}{2} + \frac{4}{3} \left( \frac{\xi}{\xi + 1} \right)^2 \right) \log(\xi + 1) \right) +  \mco(\e^{3/2}) \ , \\
F^-(\xi) &= 1 - \left( \frac{\xi}{\xi + 1} \right)^2 - \frac{\e}{3} \left( \frac{7}{6} \left( 1 - \frac{\xi}{\xi + 1} \right) \frac{\xi}{\xi + 1} - \frac{5}{6} \left( \frac{\xi}{\xi + 1} \right)^2 \log(\xi) + \rig \\
\eq \lef - \left( 1 - \frac{11}{6} \left( \frac{\xi}{\xi + 1} \right)^2 \right) \log(\xi + 1) \right)+  \mco(\e^{3/2})\ .
\end{aligned}
\end{equation}

%\newpage

%%%%%%%%%%%%%%%%%%%%%%%%%%%%%%%%%%%%%%%%%%%%%%%%%%%%%%%%%%%%
\section{An interface CFT in $6 - \e$ dimensions} \label{Sec: 6d ICFT}

In this section we bootstrap an ICFT in $6 - \e$ dimensions with cubic interactions on both sides of it. We will consider the same b.c.'s \eqref{Boundary conditions} as in section \ref{Sec: Free ICFT} (without the $O(N)$-indices), giving us the free theory correlator \eqref{Corr 2} and the free theory decomposition \eqref{Free scaling dim} and \eqref{Free Thy Decomp}. 
%Note that  only the identity appears in the bulk-channel.%, we do not need to consider a $\ph$-exchange in the free theory 
%(as in section \ref{6d BCFT}).

We can proceed to bootstrap this correlator at order $\sqrt{\e}$, assuming that the bulk anomalous dimensions does not receive corrections at this order\footnote{This assumption is motivated by the trivial CFT data one finds at order $\sqrt{\e}$ in the BCFT case.}
\begin{equation}
\begin{aligned}
\D_\ph &= \D_\ph^{(free)} + \e\g_\ph^{(1)} + \mco(\e^{3/2}) \ , \\
\D_{n\geq 1} &= 2 \left( \D_\ph^{(free)} + n \right) + \mco(\e)\ , \\
\hD_{m\geq 2} \equiv \D_{\pa_\perp^m\hp} &= \D_\ph^{(free)} + m + \sqrt{\e}\hg_m^{(1/2)} + \mco(\e) \ , \\
\end{aligned}
\end{equation}

\begin{equation}
\begin{aligned}
\la a_\1 &= 1 \ , \\
\la a_\ph &= \e^{3/2}\la a_\ph^{(3/2)} + \mco(\e^2) \ , \\
\la a_{\D_n} &= \sqrt{\e}\la a_n^{(1/2)} + \mco(\e) \ , \\
\end{aligned}
\end{equation}

\begin{equation}
\begin{aligned}
\m_\1^2 &= 0 \ , \\ 
\m^2_{\hD_m} &= \del_{m, 0} + \frac{\D_\ph^{(free)}}{2}\del_{m, 1} + \sqrt{\e}\m_m^{(1/2)} + \mco(\e) \ .
\end{aligned}
\end{equation}
Note that due to the pole in the $\ph$-block, one has to consider higher orders in its corresponding CFT data. The bootstrap calculations are very similar to those in section \ref{6d BCFT}.\footnote{The integrals and resummations are exactly the same.} We are able to express all of the CFT data in terms of one interface anomalous dimension
\begin{equation}
\begin{aligned}
\hg_1^{(1/2)} &= 2\hg_0^{(1/2)} \ , \\
\end{aligned}
\end{equation}

\begin{equation}
\begin{aligned}
\la a_\ph^{(3/2)} &= \frac{9}{4}\g_\ph^{(1)}\hg_0^{(1/2)} = -\frac{\hg_0^{(1/2)}}{8} \ , \\ 
\la a_1^{(1/2)} &= 0 \ , \\
\la a_{n\geq 2}^{(1/2)} &= - \frac{(-1)^n(n - 1)!(n + 1)!}{(2n)!}\frac{\hg_0^{(1/2)}}{2} \ ,
\end{aligned}
\end{equation}
\begin{equation}
\begin{aligned}
\m_0^{(1/2)} &= 0 \ , \\ 
\m_1^{(1/2)} &= \hg_0^{(1/2)} \ , \\ 
\m_2^{(1/2)} &= -\frac{3\hg_0^{(1/2)}}{4} \ , \\
\m_{m \geq 3}^{(1/2)} &= -\frac{ \sqrt{\pi}(-1)^m(m + 1)\G_{m - 1} }{ 2^{2m - 1}\G_{m - 1/2} }\hg_0^{(1/2)} \ .
\end{aligned}
\end{equation}
The constraints on $\hg_1^{(1/2)}$ and $\la a_\ph^{(3/2)}$ are found by removing branch cuts in $H_i$ as well as resumming the boundary-channel and comparing it with the original $H_i$ respectively. The full correlator is given by
\begin{equation*}
\begin{aligned}
F(\xi) &= 1 + \frac{ \sqrt{\e}\hg_0^{(1/2)} }{2} \left( 3\frac{\xi}{\xi + 1} \left( 1 - \frac{\xi}{\xi + 1} \right) - \left( 3 - \left( \frac{\xi}{\xi + 1} \right)^2 \right) \log(\xi + 1) \right)   + \mco(\e)\ .
\end{aligned}
\end{equation*}
For the RG domain wall near six dimensions (when one side of the interface is free), there are no corrections at order $\sqrt{\e}$. In principle one could proceed to bootstrap at order $\e$, although we are only able to commute the discontinuities with the series' in the boundary-channel in the case when the theory is completely free. It seems like more delicate methods are required in order to bootstrap this theory at order $\e$.\footnote{In the case when both sides are interacting and $\hg_0^{(1/2)}$ is zero, i.e. when there are no corrections at order $\sqrt{\e}$, one can use the method in section \ref{models} to bootstrap the theory at order $\e$. In such case, removing the branch cuts in $H_i$ allows $\hg_1^{(1)}$ to be expressed in terms of $\g_\ph^{(1)}$ and $\hg_0^{(1)}$.}

%%%%%%%%%%%%%%%%%%%%%%%%%%%%%%%%%%%%%%%%%%%%%%%%%%%%%%%%%%%%%%%%%%%%%%%%%%%%%%%%%%
\section{Conclusions}\label{Conc}

In this paper we studied a CFT with an interface or a boundary. Exploiting the analytic properties of the bulk and boundary conformal blocks we have shown how to extract the CFT data from the bootstrap equation obtained from the two-point correlation function of bulk scalar operators. This is illustrated in the context of perturbative WF theory. We have shown that the bootstrap equation upto the leading order in the $\e$-expansion contains only a few conformal blocks in either channel. This allows us to compute the OPE coefficients at the next order in epsilon. The primary model we studied is the ICFT near four dimensions with quartic interactions where we fixed the OPE coefficients in terms of the  anomalous dimensions of the operators in the spectrum. We computed the two-point correlator $\langle \f \f \rangle$ upto $\mco(\e^2)$.  We have shown how to constrain the CFT data further by using the e.o.m. on the two-point correlator which is summarised in \ref{CFTdatasummary}. This is discussed in the context of RG domain wall when one side of the interface is free. We also studied the CFT near six dimensions with cubic interactions in the presence of a boundary with Dirichlet/Neumann b.c., or an interface, where we  fixed the coefficients as well as the correlator upto $\mco(\e)$ or $\mco(\sqrt{\e})$ respectively.  At higher orders in $\e$ there are infinite number of new operators in the bootstrap equation. Hence there is a possibility of having degenerate operators in both the channels which implies that the OPE coefficients can contain contributions from multiple operators. Then one needs to solve the mixing problem and disentangle the operators to go to higher orders in $\e$.

There are several future directions that one can pursue. It would be interesting to apply this approach to compute correlators of composite correlators. For example one can study $\langle \f^2 \f^2 \rangle$ in the $\e$-expansion. This is known upto $\mco(\e)$ \cite{McAvity:1995zd}, and already in the free theory its block decompositions consists of infinitely many operators in both channels \cite{Liendo:2012hy}. It would therefore be interesting to develop bootstrap techniques to compute this correlator at the next order.

There are other theories where one can apply this method. One can consider $O(N)$-vector models at large $N$ using non-linear $\sigma$-model, or study theories with fermions, e.g. the large $N$ Gross-Neveu model or the Gross-Neveu-Yukawa model.
One can also generalize the techniques of this paper to lower dimensional defects. 

It would be interesting to study the two-point function of spinning operators in a CFT with a boundary or defect \cite{Lauria:2017wav, Billo:2016cpy, Lauria:2018klo}. For example, one can study the bootstrap techniques to the current and stress-energy two-point functions. Their block decompositions were studied in \cite{Herzog:2017xha, Herzog:2020bqw}. We hope to report on this in future.

%%%%%%%%%%%%%%%%%%%%%%%%%%%%%%%%%%%%%%%%%%%%%%%%%%%%%%%%%%%%%%%%%%%%%%%%%%%%%%%%%%

%%%%%%%%%%%%%%%%%%%%%%%%%%
\section*{Acknowledgments}
%%%%%%%%%%%%%%%%%%%%%%%%%%

We thank Agnese Bissi, Hans Werner Diehl, Pedro Liendo, Marco Meineri and Emilio Trevisani for useful discussions. We also thank Agnese Bissi, Tobias Hansen and Marco Meineri for comments on the draft. This research received funding from the Knut and Alice Wallenberg Foundation grant KAW 2016.0129 and the VR grant 2018-04438.

\appendix

\section{Boundary conditions in an ICFT} \label{App: BC}

% In this appendix we  discuss the b.c.'s in an ICFT, and motivate our choice of b.c.'s at \eqref{Boundary conditions}.
In this appendix we motivate our choice of b.c.'s at \eqref{Boundary conditions} used in the ICFT we consider. This discussion will be a generalization of the b.c.'s in two-dimensions \cite{Bachas:2001vj}, and it was briefly discussed in \cite{Herzog:2017xha}. 

On the interface, there will be a pseudo stress-energy (SE) tensor $\hta^{ab}$ with \\ 
$a, b \in \{1, ..., d - 1\}$. Its parallel derivative $\pa_\para^b\hta^{ba}$ is the interface limit of the $T^{a\perp}$-component of the bulk SE tensor. This operator measures the energy emitted or absorbed by the interface. Since there is one bulk SE tensor on each side of the interface: $T^{\m\n}_+$ and $T^{\m\n}_-$ with \\
$\m, \n \in \{1, ..., d\}$, we identify them on the interface, giving us the following b.c.
\begin{equation} \label{General BC}
\begin{aligned}
\pa_\para^b\hta^{ba} &= \hT^{a\perp}_+ = \hT^{a\perp}_- \ .
\end{aligned}
\end{equation}
In case of a fundamental scalar \eqref{Free scaling dim}, the $T^{a\perp}_\pm$-component of the SE tensor (with correction) is given by\footnote{This can be seen by varying the action w.r.t. the metric, see e.g. \cite{Prochazka:2019fah}.}
\begin{equation}
\begin{aligned}
T^{a\perp}_\pm = \pa^a\ph_\pm^{i_\pm}\pa^\perp\ph_\pm^{i_\pm} - \z\pa^a\pa^\perp(\ph_\pm^{i_\pm})^2 + \mco(\la) \ , \quad \z = \frac{d - 2}{4(d - 1)} \ , \quad i_\pm \in \{1, ..., N_\pm\} \ .
\end{aligned}
\end{equation}
Here we let the scalars on the two sides of the interface have different amount of flavours, with $N_+ \geq N_-$, and the $\mco(\la)$-terms are corrections from bulk interactions. In the interface limit we also pick up corrections from interface interactions $\hat{\la}$
\begin{equation}
\begin{aligned}
\lim\limits_{z\rightarrow 0}T^{a\perp}_\pm = \lim\limits_{z\rightarrow 0}(\pa^a\ph_\pm^{i_\pm}\pa^\perp\ph_\pm^{i_\pm}) - \z\pa^a\lim\limits_{z\rightarrow 0}(\pa^\perp(\ph_\pm^{i_\pm})^2) + \mco(\la, \hat{\la}) \ .
\end{aligned}
\end{equation}
Given this, one solution to the b.c. equation \eqref{General BC} is to relate the fundamental field and its normal derivative for the first $N_-$ scalars at the interface, and consider \textbf{reflective b.c.'s} for the excessive scalars (i.e. either Dirichlet and Neumann b.c.'s similar to those in a BCFT)\footnote{One can see this by studying the boundary limit of $T^{a\perp}_\pm$ inserted into a correlator, e.g. $\langle T^{a\perp}_\pm(x)\hat{O}^2(y)\rangle$, where $\hat{O}^2$ is a composite operator on the interface.}
\begin{equation}
\begin{aligned}
\lim\limits_{z\rightarrow 0}\ph_+^i &= \lim\limits_{z\rightarrow 0}\ph_-^i \ , \quad &\lim\limits_{z\rightarrow 0}\pa_\perp\ph_+^i &= \lim\limits_{z\rightarrow 0}\pa_\perp\ph_-^i \ , \quad &i&\in\{1, ..., N_-\} \ , \\
\lim\limits_{z\rightarrow 0}\ph_+^m &= 0 \ , \quad \text{or ,} \quad &\lim\limits_{z\rightarrow 0}\pa_\perp\ph_+^m &= 0 \ , \quad &m&\in\{1, ..., N_+ - N_-\} \ .
\end{aligned}
\end{equation}
If we have $N_+ > N_-$ in above b.c.'s, we have reduced the $O(N_+) \times O(N_-)$ ICFT that we started studying into a $O(N_-) \times O(N_-)$ ICFT and a $O(N_+ - N_-)$ BCFT (assuming there are no interactions that mix these two systems). These are two systems that one can study separately. Bootstrap methods in BCFTs with $O(N)$ symmetry was studied in \cite{Bissi:2018mcq}. In this paper we study bootstrap methods in the other system, i.e. an ICFT with $O(N)\times O(N)$ symmetry. Hence we consider the amount of flavours to be the same on both sides of the interface in section \ref{Sec: Free ICFT}. In such case, we have the b.c.'s at \eqref{Boundary conditions}.

\section{The $\e$-expansion of boundary conformal blocks} \label{App: Block Exp}

Here we explain how we $\e$-expand the two boundary blocks $\mG_{\rm{boe}}(\hD_0, \xi)$ and $\mG_{\rm{boe}}(\hD_1, \xi)$ using the algorithm in \cite{Huber:2005yg}. The main idea is to use the following integral representation for the hypergeometric functions, and $\e$-expand its integrand before performing the integration
\begin{equation} \label{Int Repr}
\begin{aligned}
{}_2F_1(a, b, c, z) = \frac{\G_c}{\G_b\G_{c - b}}\int_{0}^{1}dt\frac{t^{b - 1}(1 - t)^{c - b - 1}}{(1 - tz)^a} \ .
\end{aligned}
\end{equation}
However, the $\e$-expansion and the integral only commutes when the integral converges, which occurs when $c|_{\e = 0} > b|_{\e = 0} > 0$. To guarantee this we will use the following recurrence relations for the hypergeometric function, and use the above integral representation for each term \\
\begin{equation} %\label{2F1 Rel 1} %\label{2F1 Rel 2}
\begin{aligned} 
\, _2F_1(a,b;c;z) &= -\frac{z (a + b - 2c - 1) + c }{ c(z - 1) }\, _2F_1(a,b;c+1;z) + \\
\eq - \frac{ z (a - c - 1) (b - c - 1) }{ c(c + 1)(z - 1) }\, _2F_1(a,b;c+2;z) \ .
%\frac{z (-a-b+2 c+1)-c}{c (z-1)}\, _2F_1(a,b;c+1;z) + \frac{z (a-c-1) (-b+c+1)}{c (c+1) (z-1)}\, _2F_1(a,b;c+2;z) \ .
%\, _2F_1(a,b;c;z) &= \frac{b-c-1}{(b-1) (z-1)}\, _2F_1(a,b-2;c;z) + \\
%\eq + \frac{z (-a+b-1)-2 b+c+2}{(b-1) (z-1)}\, _2F_1(a,b-1;c;z) \ , \\
\end{aligned}
\end{equation}
After expanding in $\e$, we need to integrate over $t$. To perform this integration, we use the Mathematica command HypExpInt \cite{Huber:2005yg}, which evaluates the integral
\begin{equation}
\begin{aligned}
\text{HypExpInt}(a_1, a_2, a_3, a_4 , a_5, z) &= \int_{0}^{1}dt\frac{t^{a_1}\log(t)^{a_2}\log(1 - t)^{a_3}\log(1 - tz)^{a_4}}{(tz - 1)^{a_5}} \ .
\end{aligned}
\end{equation}
This yields the following expansions of hypergeometric functions in $\mG_{\rm{boe}}(\hD_0, \xi)$ and $\mG_{\rm{boe}}(\hD_1, \xi)$
\begin{equation*}
\begin{aligned}
&\, _2F_1\left(\frac{1}{2} (-2 \alpha -1) \epsilon + \hg_0^{(2)} \epsilon ^2+1,\hg_0^{(2)} \epsilon ^2-\alpha  \epsilon ;2 \hg_0^{(2)} \epsilon ^2-2 \alpha  \epsilon ;z\right) = \\
\eq = \frac{z-2}{2 (z-1)} + \epsilon  \left(\frac{\alpha }{2}+\frac{1}{4-4 z}\right) \log
(1-z) + \epsilon ^2 \left(\frac{(1-2 \alpha ) \alpha  z \text{Li}_2(z)}{4 (z-1)} + \rig \\
\tq \lef +\left(\frac{\alpha }{8}+\frac{1}{16-16 z}\right) \log ^2(1-z) -\frac{1}{2} \hg_0^{(2)} \log (1-z)\right) \ , \\
\end{aligned}
\end{equation*}
\begin{equation*}
\begin{aligned}
&{}_2F_1\left(2 - \left( \frac{1}{2} + \al \right) \e + \hg_1^{(2)}\e^2, 1 - \al\e + \hg_1^{(2)}\e^2, 2 - 2\al\e + 2\hg_1^{(2)}\e^2, z\right) = \\
\eq = \frac{1}{1-z} + \epsilon  \left(\frac{\alpha -\frac{1}{2}}{z-1}+\frac{\left(\alpha +\alpha  (-z)-\frac{1}{2}\right) \log (1-z)}{(z-1) z}\right) + \\
\tq + \epsilon ^2 \left(\frac{\left(\alpha -\frac{1}{2}\right) \alpha  (z-2) \text{Li}_2(z)}{(z-1) z}+\frac{(-2 \alpha +2 \alpha  z+1) \log ^2(1-z)}{8 z-8 z^2} + \rig \\
\tq \lef + \frac{-4 \alpha ^2+4 \hg_1^{(2)}+1}{4-4 z}+\frac{\log (1-z) \left(4 \alpha ^2 (z-1)-2 \alpha 
	(z-2)+4 \hg_1^{(2)} (z-1)-1\right)}{4 (z-1) z}\right) \ .
\end{aligned}
\end{equation*}

\section{Resummation techniques}

In this appendix we describe the different techniques we use for resuming the bulk and boundary-channels (using Mathematica).

\subsection{Expansion in $\xi$} \label{App: Bulk Resum}

We use this technique when resuming the bulk-channel in \eqref{Bootstrap eq}.

\begin{enumerate}
	\item Use first the integral representation \eqref{Int Repr} for the hypergeometric functions.
	\item Resum the blocks.
	\item Expand the integrand around $\xi = 0$. This means that after we have performed the integration, we have to resum the expansion in $\xi$. For each power in $\xi$ we have polynomials in the integration parameter. 
	\item Integrate the polynomials.
	\item Find a general term in the expansion of $\xi$.
	\item Resum the expansion in $\xi$. This is the final result.
\end{enumerate}

\subsection{Differentiate with respect to $\xi$ and simplify using an ansatz} \label{App: Boundary Resum}

This method is used when resuming even and odd operators in the boundary-channel in \eqref{Bootstrap eq}.

\begin{enumerate}
	\item Use first the integral representation \eqref{Int Repr} for the hypergeometric functions.
	\item \label{2C} Resum the blocks.
	\item For odd blocks ($\hD = 2m + 1 \ , m\in\Z_{\geq 0}$), differentiate with respect to $\xi$. This is only useful when the derivative is on a simpler form as the original integrand, e.g. when taking the derivative removes dilogarithmic terms.
	\item Perform the integration from \eqref{Int Repr}.
	\item If a derivative w.r.t. $\xi$ was taken, perform an indefinite integrate over $\xi$. This will yield an integration constant that we need to fix.
	\item The integration constant is fixed by expanding our result around large $\xi \gg 1$. Compare this with the expansion of the original integrand in step \ref{2C}, where we only integrate the term at lowest order ($\xi^0$).
	\item This is the final result. Near four dimensions it may be a cumbersome expression that contains several different dilogarithms. One way to simplify it is to make the following ansatz (with the correct branch cuts only along $\xi \in (-1, 0)$) for your result
	\begin{equation} \label{Ansatz}
	\begin{aligned}
	f(\xi) &= A_1 + A_2\frac{\xi}{\xi + 1} + \left( A_3 + A_4\frac{\xi}{\xi + 1} \right) \log\left( \frac{\xi}{\xi + 1} \right) + \\
	\eq + \left( A_5 + A_6\frac{\xi}{\xi + 1} \right) \log^2\left( \frac{\xi}{\xi + 1} \right) + \left( A_7 + A_8\frac{\xi}{\xi + 1} \right) \Li_2\left( -\frac{1}{\xi} \right) \ .
	\end{aligned}
	\end{equation}
	\item The coefficients $A_i \ , i\in\{1, ..., 8\}$ in our ansatz can be found by comparing the expansions around large $\xi$ with that of the final result from the previous step. %We noted that several of the dilogarithms in the final result requires a long time to be expanded using Mathematica (one can check which of these dilogarithms are problematic by expanding each one of them on their own). 
	Some of the  dilogarithms can be expanded after using the following identity
	\begin{equation}
	\begin{aligned}
	\Li_2(z) &= \frac{\pi^2}{6} - \Li_2(1 - z) - \log(z)\log(1 - z) \ .
	\end{aligned}
	\end{equation}
	%\item To make sure that the ansatz \eqref{Ansatz} (with the coefficients fixed through step 8) agree with the final result, we plot the two over a large range of $\xi > 0$ to make sure that they are the same.
\end{enumerate}

\bibliographystyle{utphys}
\footnotesize
\bibliography{paper}

\end{document}